\documentclass{agujournal2019}
\usepackage{url} %this package should fix any errors with URLs in refs.
\usepackage{lineno}
\usepackage[inline]{trackchanges} %for better track changes. finalnew option will compile document with changes incorporated.
\usepackage{soul}
\usepackage{amsmath}
\journalname{Journal of Advances in Modeling Earth Systems (JAMES)}

\begin{document}

%%%%%%%%%%%%%%%%%%%%%%%%%%%%%%%%%%%%%%%%%%%%%%%
%  TITLE
%
% (A title should be specific, informative, and brief. Use
% abbreviations only if they are defined in the abstract. Titles that
% start with general keywords then specific terms are optimized in
% searches)
%
%%%%%%%%%%%%%%%%%%%%%%%%%%%%%%%%%%%%%%%%%%%%%%%

% Example: \title{This is a test title}

\title{Ocean Wave Forecasting with Deep Learning as Alternative to Conventional Models}

%%%%%%%%%%%%%%%%%%%%%%%%%%%%%%%%%%%%%%%%%%%%%%%
%
%  AUTHORS AND AFFILIATIONS
%
%%%%%%%%%%%%%%%%%%%%%%%%%%%%%%%%%%%%%%%%%%%%%%%

% Authors are individuals who have significantly contributed to the
% research and preparation of the article. Group authors are allowed, if
% each author in the group is separately identified in an appendix.)

% List authors by first name or initial followed by last name and
% separated by commas. Use \affil{} to number affiliations, and
% \thanks{} for author notes.
% Additional author notes should be indicated with \thanks{} (for
% example, for current addresses).

% Example: \authors{A. B. Author\affil{1}\thanks{Current address, Antartica}, B. C. Author\affil{2,3}, and D. E.
% Author\affil{3,4}\thanks{Also funded by Monsanto.}}

\authors{Ziliang Zhang\affil{1,2}\thanks{These authors contributed equally to this work.}, Huaming Yu\affil{1,2,3}\thanks{These authors contributed equally to this work.}, Danqin Ren\affil{4}, Chenyu Zhang\affil{5}, Minghua Sun\affil{6}, Xin Qi\affil{7}.}
% Author\affil{3,4}\thanks{Also funded by Monsanto.}}

% \affiliation{1}{First Affiliation}
% \affiliation{2}{Second Affiliation}
% \affiliation{3}{Third Affiliation}
% \affiliation{4}{Fourth Affiliation}

\affiliation{1}{College of Oceanic and Atmospheric Sciences, Ocean University of China, Qingdao, China}
\affiliation{2}{State Key Laboratory of Physical Oceanography, Ocean University of China, Qingdao, China}
\affiliation{3}{Sanya Oceanographic Institution, Ocean University of China, Sanya, China}
\affiliation{4}{Dawning Information Industry Co., Ltd, Qingdao, China}
\affiliation{5}{Qingdao Ekman Technology Co., Ltd, Qingdao, China}
\affiliation{6}{CMA Earth System Modeling and Prediction Centre (CEMC), China Meteorological Administration, Beijing, China}
\affiliation{7}{College of Management, Ocean University of China, Qingdao, China}
%(repeat as many times as is necessary)

% Corresponding author mailing address and e-mail address:

% (include name and email addresses of the corresponding author.  More
% than one corresponding author is allowed in this LaTeX file and for
% publication; but only one corresponding author is allowed in our
% editorial system.)

% Example: \correspondingauthor{First and Last Name}{email@address.edu}

\correspondingauthor{Huaming Yu}{hmyu@ouc.edu.cn}
\correspondingauthor{Xin Qi}{x.qi@ouc.edu.cn}

%%%%%%%%%%%%%%%%%%%%%%%%%%%%%%%%%%%%%%%%%%%%%%%
% KEY POINTS
%%%%%%%%%%%%%%%%%%%%%%%%%%%%%%%%%%%%%%%%%%%%%%%
%  List up to three key points (at least one is required)
%  Key Points summarize the main points and conclusions of the article
%  Each must be 140 characters or fewer with no special characters or punctuation and must be complete sentences

% Example:
% \begin{keypoints}
% \item	List up to three key points (at least one is required)
% \item	Key Points summarize the main points and conclusions of the article
% \item	Each must be 140 characters or fewer with no special characters or punctuation and must be complete sentences
% \end{keypoints}

\begin{keypoints}
\item A deep learning wave model shows forecast skill comparable to operational systems with high computational efficiency
\item An auto-regressive process with wind forcing enables forecasting of waves during extreme weather events
\item Satellite validation shows regional model strengths with the new model excelling in the West Pacific
\end{keypoints}

%%%%%%%%%%%%%%%%%%%%%%%%%%%%%%%%%%%%%%%%%%%%%%%
%
%  ABSTRACT and PLAIN LANGUAGE SUMMARY
%
% A good Abstract will begin with a short description of the problem
% being addressed, briefly describe the new data or analyses, then
% briefly states the main conclusion(s) and how they are supported and
% uncertainties.

% The Plain Language Summary should be written for a broad audience,
% including journalists and the science-interested public, that will not have 
% a background in your field.
%
% A Plain Language Summary is required in GRL, JGR: Planets, JGR: Biogeosciences,
% JGR: Oceans, G-Cubed, Reviews of Geophysics, and JAMES.
% see http://sharingscience.agu.org/creating-plain-language-summary/)
%
%%%%%%%%%%%%%%%%%%%%%%%%%%%%%%%%%%%%%%%%%%%%%%%

%% \begin{abstract} starts the second page

\begin{abstract}
This study presents OceanCastNet (OCN), a machine learning approach for wave forecasting that incorporates wind and wave fields to predict significant wave height, mean wave period, and mean wave direction.We evaluate OCN's performance against the operational ECWAM model using two independent datasets: NDBC buoy and Jason-3 satellite observations. NDBC station validation indicates OCN performs better at 24 stations compared to ECWAM's 10 stations, and Jason-3 satellite validation confirms similar accuracy across 228-hour forecasts. OCN successfully captures wave patterns during extreme weather conditions, demonstrated through Typhoon Goni with prediction errors typically within ±0.5 m. The approach also offers computational efficiency advantages. The results suggest that machine learning approaches can achieve performance comparable to conventional wave forecasting systems for operational wave prediction applications.
\end{abstract}

\section*{Plain Language Summary}
Predicting ocean waves is important for maritime safety and coastal planning, but current computer models require powerful computers and take many hours to run. This study developed OceanCastNet, a machine learning model that learns to predict wave height, period, and direction by studying patterns in historical wave and wind data. We tested our approach against ECWAM, a widely-used operational wave forecasting system, using two different types of validation data: ocean buoy measurements, and satellite observations. The results show that our machine learning approach can predict waves about as accurately as conventional models, performing better at 24 buoy locations compared to the conventional model's better performance at 10 locations, and successfully capturing wave patterns during Typhoon Goni in 2020. The machine learning approach also runs much faster, completing forecasts in seconds rather than hours, suggesting that artificial intelligence offers a promising alternative for wave forecasting applications.

%%%%%%%%%%%%%%%%%%%%%%%%%%%%%%%%%%%%%%%%%%%%%%%
%
%  BODY TEXT
%
%%%%%%%%%%%%%%%%%%%%%%%%%%%%%%%%%%%%%%%%%%%%%%%

\section{Introduction}
Traditional geophysical forecasting models are fundamentally grounded in physical conservation principles and the integration of external forcings \cite{mellorUsersGuideThree1998,skamarockDescriptionAdvancedResearch2008}. In wave science, leading systems like the widely trusted ECWAM conserve momentum and mass \cite{group1988wam} by discretizing governing equations on high-resolution grids. While physically robust, this pursuit of accuracy through grid refinement leads to substantial computational demands, posing a significant challenge for real-time operational forecasting.

In parallel, deep learning (DL) has emerged as a transformative approach in meteorology. Models like FourCastNet, Pangu-Weather, and GraphCast have demonstrated remarkable success in atmospheric forecasting, with accuracies improving from being comparable \cite{pathakFourcastnetGlobalDatadriven2022} to outperforming traditional methods in specific metrics such as RMSE \cite{biPanguweather3dHighresolution2022,lamGraphCastLearningSkillful2022}. These models deliver predictions at speeds far exceeding traditional methods, though challenges for operational integration remain, particularly in data assimilation systems \cite{slivinski2025assimilating,tian2024exploring}. However, the application of DL to global wave forecasting has been less developed. Existing DL-based wave models are often limited to single-point or regional predictions \cite{minuzziDeepLearningApproach2023,adytiaDeepLearningApproach2022,zhengMultivariateDataDecomposition2023,bai2022development,jing2022numerical}, lack comprehensive outputs like wave direction and period, or have insufficient temporal resolution, as seen in models like AI-GOMS \cite{xiongAIGOMSLargeAIDriven2023}. This hinders their adoption in operational settings that demand global coverage and detailed, frequent updates.

To bridge this gap, this paper introduces OceanCastNet (OCN), a global deep learning wave forecasting model designed as a computationally efficient and operationally viable alternative to traditional systems. Our approach demonstrates performance comparable to the benchmark ECWAM model across multiple validation datasets while achieving a significant computational speedup. This work establishes the potential of deep learning to provide accurate, fast, and comprehensive global wave forecasts suitable for operational use.

\section{Data and Methods}
\subsection{Data Sources and Preprocessing}

For model training, we use ERA5 reanalysis data \cite{hersbachERA5HourlyData2023} from the European Centre for Medium-Range Weather Forecasts (ECMWF). ERA5 provides global hourly 10-m wind field data at 0.25-degree resolution and hourly significant wave height, mean wave period, and mean wave direction data at 0.5-degree resolution. These data integrate multi-source observation data including conventional observations, satellite scatterometers, and satellite altimeters \cite{hersbachERA5GlobalReanalysis2020}. It is worth noting that while significant wave height is well-constrained by satellite observations, mean wave period and direction are of moderate quality due to limited observational constraints. We use 6-hour temporal resolution and select the latitude range from -79.5° to 80° to maintain consistency with operational forecasting requirements. We use 41 years of ERA5 data from 1980 to 2020, with 1980-2017 as the training set, 2018-2019 as the validation set, and 2020 as the test set.

For model performance evaluation as an alternative to conventional wave models, we collected data from multiple sources. ECWAM operational forecast data for 2024 was used for comparison. ECWAM serves as the operational wave forecasting component of ECMWF's Integrated Forecasting System and represents the standard for operational wave prediction systems. ECWAM's native resolution of 0.25° was resampled to 0.5° to match our model's output grid for direct comparison. To ensure equivalent input conditions between OCN and conventional wave models, we obtained IFS wind forecast data for 2024 through the TIGGE (The International Grand Global Ensemble) database \cite{tiggeDataset2010}.

To evaluate the model's performance at specific locations where conventional wave models are routinely validated, we use buoy data from the National Data Buoy Center (NDBC) for 2020 and 2024\cite{ndbcData2024}. For additional independent validation, we acquired Jason-3 satellite altimeter observations from PO.DAAC \cite{desaiJason3GPSBased2016}, which provide satellite-based wave height measurements independent of both conventional and machine learning wave models. ERA5 data for 2024 serves as an additional reference dataset for comprehensive evaluation.

\subsection{Model Architecture and Design}

The OCN model consists of a core deep learning architecture for single-step prediction and an auto-regressive workflow for generating continuous forecasts, as illustrated in the two panels of Figure~\ref{fig:model_architecture}.

Wave fields can be characterized by two key components: wind sea, which is directly influenced by wind-sea surface interactions in wind-affected areas, and swell, which forms as wind sea propagates beyond its generation zone \cite{wiegelWindWavesSwell1960}. To model these complex processes, we propose OCN, a model designed to capture spatial-temporal evolution patterns of waves while explicitly incorporating external forcing.

OCN builds upon the Adaptive Fourier Neural Operators (AFNO) framework, which has shown excellent performance in solving partial differential equations \cite{liFourierNeuralOperator2020, guibasAdaptiveFourierNeural2021}. AFNO utilizes Fast Fourier Transform (FFT) and Inverse Fast Fourier Transform (IFFT) to enable efficient spatial and multi-channel convolution. This approach serves as a replacement for the attention mechanism in Vision Transformers (ViT), significantly reducing computational complexity from O(N²) to O(N log N). This efficiency is crucial for processing the high-resolution, multi-channel data typical in oceanographic studies.

The core architecture for a single forecast step is shown in Figure~\ref{fig:model_architecture}, panel (a). Based on the FourCastNet model, the input tensor is first divided into non-overlapping patches and linearly projected into a high-dimensional latent space, with positional encodings added to retain spatial information. These tokens are then processed by a series of 12 AFNO layers, which perform global spatial mixing via a Fourier transform-based operator and local channel mixing using an MLP. Finally, a linear decoder reconstructs the predicted wave parameters on the original grid. The full architectural and training hyperparameters are detailed in Appendix A(Table~\ref{tab:hyperparameters}).

To generate forecasts beyond a single step, OCN employs an auto-regressive workflow, shown in Figure~\ref{fig:model_architecture}, panel (b), which uses the following input-output relationship:
\begin{equation}
[W(t+1)]=f[u_{10}(t-1), v_{10}(t-1), W(t-1), u_{10}(t), v_{10}(t), W(t), u_{10}(t+1), v_{10}(t+1)]
\end{equation}
where
\begin{equation}
W(t) = [H_s(t), T_m(t), \theta_m(t)]
\end{equation}
In this equation, $H_s$, $T_m$, and $\theta_m$ represent significant wave height, mean wave period, and mean wave direction, respectively. The variables $u_{10}$ and $v_{10}$ denote the 10m u-component and v-component of wind. All variables are considered at various time steps $(t)$, where each time step represents a duration of 6 hours. The specific multi-timestep input configuration shown in Equation (1) was selected based on a series of ablation studies designed to find the optimal balance between predictive performance and computational efficiency. These studies, which evaluated various combinations of historical and future timesteps, are detailed in the Supplementary Material (Section S1).

As illustrated in Figure~\ref{fig:model_architecture}, panel (b), the OCN prediction workflow follows an iterative process. The model initiates by processing an input tensor comprising wind and wave parameters from three consecutive time steps (t-1, t, and t+1). This initial computation yields the first prediction $[H_s(t+1), T_m(t+1), \theta_m(t+1)]$. For subsequent iterations, OCN constructs a new input tensor by advancing the time window, incorporating the previous prediction as the current wave field, and introducing the next set of future wind data.

This multi-timestep, auto-regressive design is crucial for the model's performance. By integrating past wave states with current and future external wind forcing at each step, OCN can effectively learn the temporal evolution of wave dynamics. This forward-looking approach, which continuously incorporates the latest atmospheric data, ensures the forecast remains sensitive to changing weather conditions and aligns the model's methodology with the physically-driven nature of traditional wave models.

\subsection{Training Procedure and Land Masking}

\begin{figure}
  \centering
  \noindent\includegraphics[width=\textwidth]{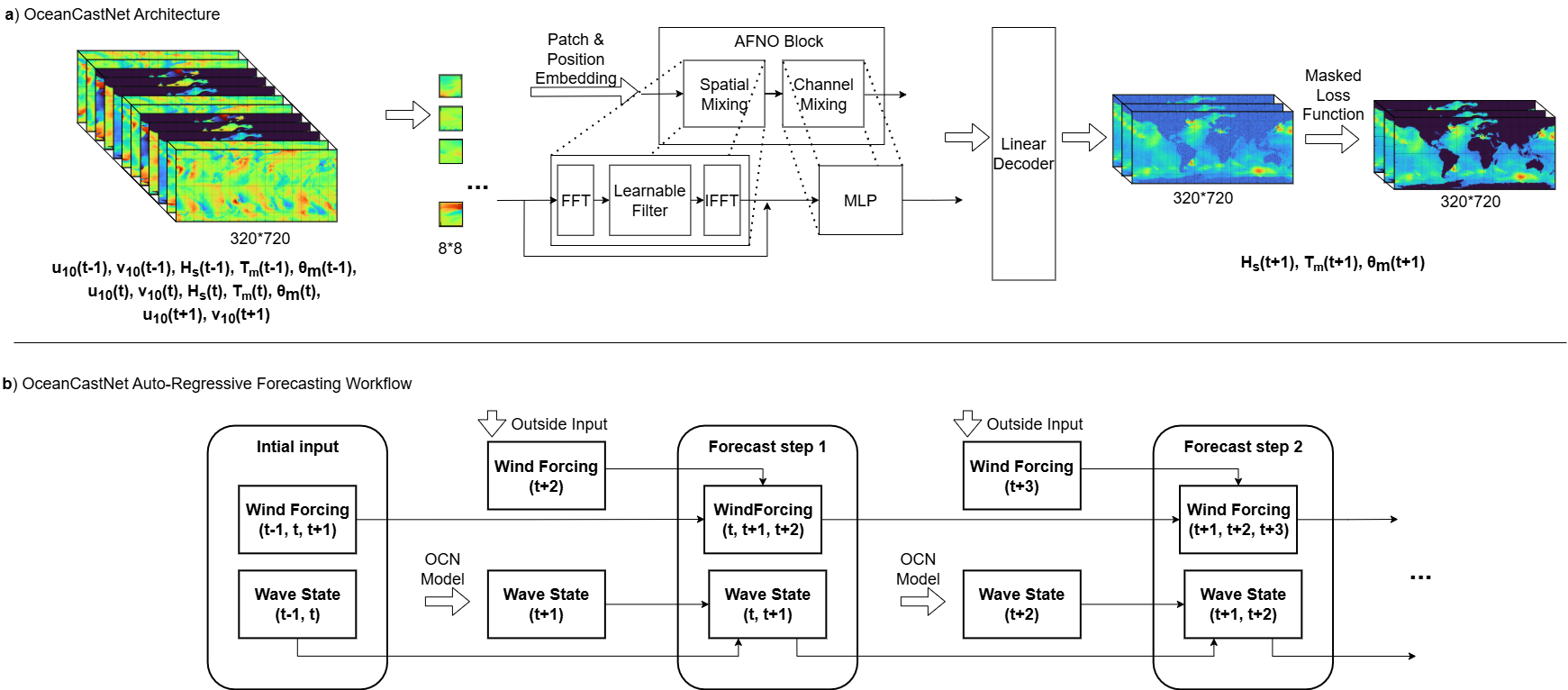}
  \caption{OceanCastNet Architecture and Auto-Regressive Workflow.
(a) Model Architecture for a single forecast step. The input tensor, containing wave and wind data from multiple time steps, is partitioned into non-overlapping patches. These patches are linearly embedded into high-dimensional tokens with added positional encoding. The tokens are then processed by a series of 12 AFNO Layers, where information is mixed globally via a Fourier transform-based spatial mixer and locally via an MLP-based channel mixer. Finally, a linear decoder reconstructs the output grid of predicted wave parameters, which is then passed to a masked loss function for error calculation during training.
(b) Auto-Regressive Forecasting Workflow. The model is initialized with a known wave state (t-1, t) and corresponding wind forcing. In "Forecast step 1," the model (using the architecture from panel (a) predicts the wave state at t+1. For "Forecast step 2," this predicted state becomes part of the new input, and the model is provided with new, externally-supplied future wind forcing (t+3) to predict the wave state at t+2. This iterative process continues for the entire forecast duration.}
  \label{fig:model_architecture}
\end{figure}

To address the impact of land presence on wave model training, we implemented a masked loss function during backpropagation, as illustrated in Figure~\ref{fig:model_architecture}, panel (a). The training process distinguishes between forward and backward propagation: during forward propagation, the model processes the entire spatial domain, including both land and ocean regions, to maintain spatial coherence and predict wave parameters according to Equation (1), while during backpropagation, we apply spatial masking to focus learning exclusively on ocean regions. The land-sea mask $M$ is defined as $M_{i,j} = 1$ if pixel $(i,j)$ is ocean and $M_{i,j} = 0$ if pixel $(i,j)$ is land. The masked relative error loss function is computed as:
\begin{equation}
L = \frac{1}{N} \sum_{n=1}^{N} \frac{\|(\hat{W}_n - W_n) \odot M\|_p}{\|W_n \odot M\|_p + \epsilon} 
\end{equation}
where $\hat{W}_n$ and $W_n$ are the predicted and target wave fields for sample $n$, $N$ is the number of samples in the training batch, $\odot$ denotes element-wise multiplication, $\|\cdot\|_p$ is the $p$-norm, and $\epsilon = 10^{-8}$ prevents division by zero. This masking strategy ensures that prediction errors over land regions do not contribute to gradient updates, allowing the model to focus its learning capacity on accurate ocean wave forecasting while maintaining spatial coherence during forward propagation. Crucially, this same land mask is also applied to the model's output at each step of the iterative forecasting process. This prevents the accumulation of noisy, unconstrained predictions over land areas from contaminating the ocean forecast in subsequent steps.

Model training is conducted on 32 NVIDIA Tesla V100 GPUs, with a training time of 16 hours.

\subsection{Evaluation Metrics}

In evaluating our model's performance, we employ a combination of area-based and point-based metrics to provide a comprehensive assessment.

For area-based evaluation against gridded datasets, we adopt the ACC, RMSE, and mean error (ME) as our primary metrics. These metrics offer complementary insights into the model's forecasting capabilities. ACC measures the correlation between the model's forecasted anomalies and the actual anomalies, providing an indication of the model's ability to capture the spatial patterns and temporal variations of wave characteristics. RMSE quantifies the overall magnitude of prediction errors, giving us a measure of forecast accuracy. ME provides the difference between predicted and observed values, offering a direct measure of prediction bias at individual points without averaging across locations.

For point-based evaluation against observational data (NDBC buoy measurements and Jason-3 satellite observations), we employ the Pearson correlation coefficient (r) and RMSE using two distinct approaches depending on the analysis objective. The first approach calculates metrics for each time step across all available observation sites, which is used to evaluate the temporal evolution of model performance (sections \ref{sec:NDBCeval} and \ref{sec:Jasoncompare}). The second approach calculates metrics for each individual station across all available time steps, which enables station-by-station performance comparison (section \ref{sec:NDBCcompare}). The Pearson correlation coefficient measures the linear correlation between predicted and observed values, providing insight into the model's ability to capture local wave dynamics, while RMSE quantifies the magnitude of prediction errors.

The detailed formulas for calculating all metrics (ACC, RMSE, ME, and both point-based r and RMSE approaches) are provided in \ref{app:Formulas}.

\section{Experiments and Results}
Our experiments are designed to comprehensively evaluate the performance and capabilities of our model under various conditions. The experimental framework encompasses two main areas of investigation, each addressing different aspects of the model's performance and potential applications.

The first part of our experiment assesses the model's optimal performance under idealized conditions. By utilizing ERA5 data for both initial conditions and atmospheric forcing, we establish an idealized forecasting scenario that represents the model's potential performance when provided with reanalysis data. This approach allows us to evaluate the model's accuracy through both area-wide and point-specific analyses. While this represents an idealized scenario using reanalysis data, it establishes the model's potential performance under optimal conditions and serves as an upper bound for comparison with operational forecasting applications where real-time atmospheric forcing must be used.

In the second part, we compare our model's performance with ECWAM operational forecasts under realistic forecasting conditions. Using ECWAM operational forecast data for 2024 as our benchmark, we evaluate OCN's performance as an alternative to the current operational wave forecasting standard. We initialize our model using ECWAM results and IFS wind forecast data for subsequent predictions. This approach simulates operational forecasting scenarios and provides a fair comparison between OCN and the operational standard. By using ECWAM initialization and IFS wind forecasts, we assess OCN's performance under realistic operational conditions, thereby evaluating its effectiveness as a potential complement to conventional wave forecasting systems.

\subsection{Performance Evaluation on ERA5 Dataset}
This part consists of two main components. In the first component, we evaluate the model's performance globally for the year 2020, using ERA5 wave data as the target result. In the second component, we evaluate the model's performance at specific points for the year 2020, using NDBC buoy data as the target result. In both cases, the model is initialized with ERA5 data and continuously driven by ERA5 wind data.

\subsubsection{Global Model Evaluation}

In this experiment, we utilized ERA5 reanalysis data as our reference dataset. To ensure robust statistical analysis, we conducted a comprehensive evaluation using 36 distinct forecast cases throughout the year. Specifically, we initiated forecasts from the 1st, 4th, and 12th day of each month, using wind and wave data from 00:00 and 06:00, along with wind data at 12:00, as initial inputs to predict wave conditions at 12:00. This constitutes the first step of our prediction process. For each of these 36 initialization times, the model was run for 60 consecutive steps (approximately 15 days) with continuous ERA5 wind data input.We evaluated the model's performance by comparing its predictions to ERA5 wave data, which served as our target result. This represents an idealized validation scenario that demonstrates the model's capability to translate wind information to wave fields under optimal conditions with reanalysis data.

\begin{figure}
  \centering
  \noindent\includegraphics[width=\textwidth]{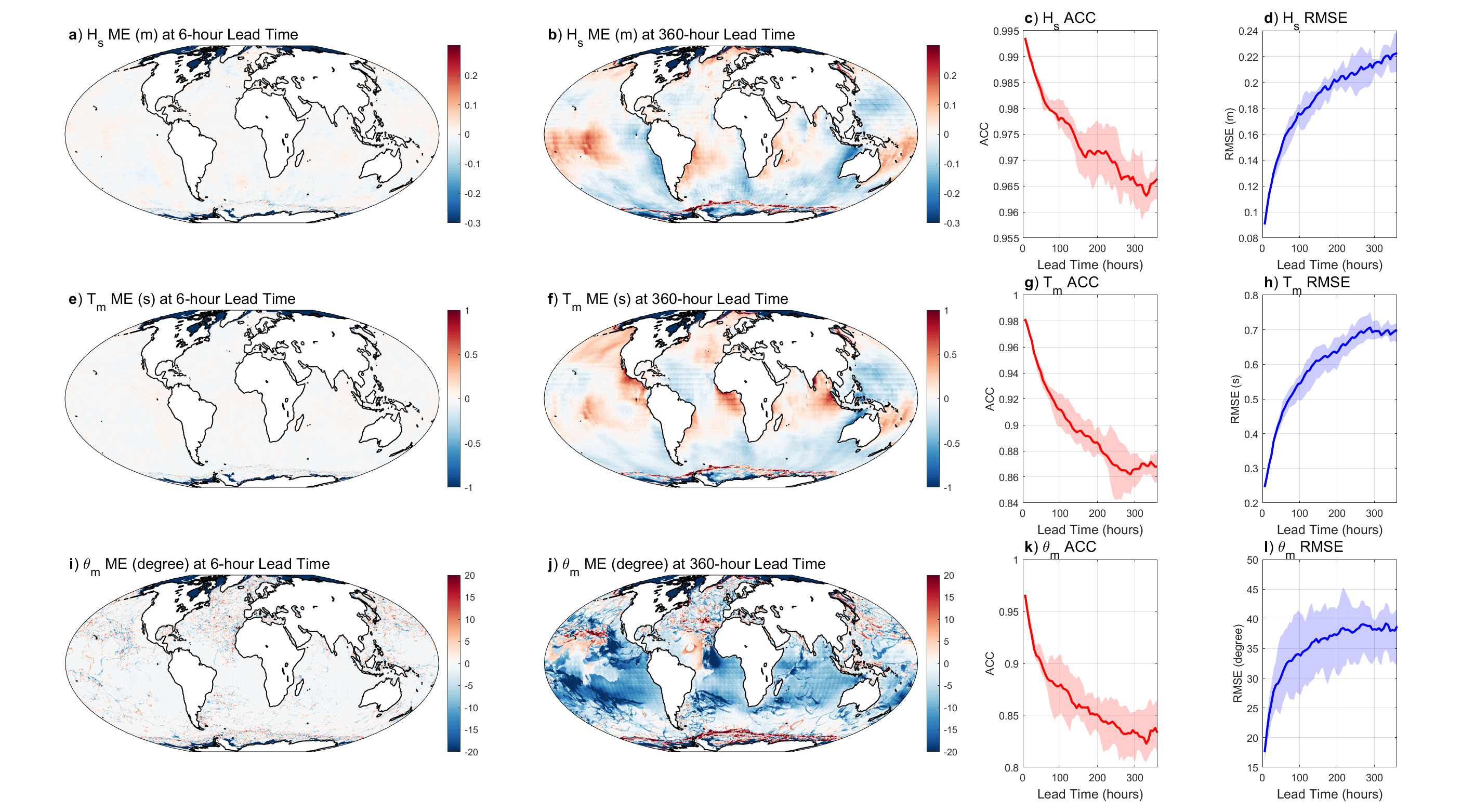}
  \caption{Global performance analysis of the wave prediction model based on 36 forecast cases throughout the year. (a-b, e-f, i-j) Spatial distribution of mean error (ME) for significant wave height ($H_s$), mean wave period ($T_m$), and mean wave direction ($\theta_m$) at 6-hour and 360-hour lead times. Red indicates positive bias (over-prediction) while blue indicates negative bias (under-prediction). (c-d, g-h, k-l) Temporal evolution of anomaly correlation coefficient (ACC) and root mean square error (RMSE) for each parameter over the 360-hour forecast period. Solid lines represent the mean across all 36 forecast cases, while shaded areas represent the interquartile range (25th to 75th percentiles).}
  \label{fig:averaged_forecast_results}
\end{figure}

Figure~\ref{fig:averaged_forecast_results} illustrates the spatial distribution of ME for different wave variables at 6-hour (first step) and 360-hour (last step) lead times (panels a, b, e, f, i, j). The ME maps with diverging colormaps reveal both the magnitude and direction of prediction biases. At the 6-hour lead time, errors are generally small and evenly distributed across positive and negative values for all three wave parameters. By the 360-hour lead time, more structured spatial patterns emerge, particularly in the Southern Ocean and northern Pacific regions, though the magnitude of ME for significant wave height and mean wave period remains below 0.3 m and 1 s respectively in most regions, while the ME for mean wave direction generally stays within ±20 degrees.

The temporal evolution of ACC and RMSE is depicted in panels (c, d, g, h, k, l) of Figure~\ref{fig:averaged_forecast_results}. Solid lines represent mean values across all 36 forecast cases, with shaded areas indicating the interquartile range (25th to 75th percentiles). The relatively narrow width of these shaded regions demonstrates the model's stability across different initialization times throughout the year. Despite gradual error accumulation over time, the mean ACC for significant wave height remains above 0.96, with RMSE below 0.25 m throughout the 360-hour forecast period. For mean wave period, the ACC stays above 0.85, with RMSE under 0.8 s.

The representation of wave direction poses some challenges for the model. The abrupt transition from 360 degrees to 0 degrees may introduce artificial discontinuities, potentially impacting model performance. Consequently, mean wave direction predictions show somewhat lower performance compared to significant wave height and mean wave period. Nevertheless, they still achieve a mean ACC above 0.8 and a mean RMSE below 40 degrees, indicating viable forecasting skill even for this more challenging parameter.

The lower performance for wave direction may partly reflect the inherent uncertainty in ERA5's wave direction data, which is not directly constrained by observations but derived from the underlying wave model. This limitation in the reference dataset should be considered when interpreting wave direction performance metrics.

To assess the model's robustness under extreme conditions, we examined its performance during a significant weather event of 2020. For this analysis, we selected Typhoon Goni, the strongest typhoon of 2020 in the Western Pacific, as our case study. Goni was classified as a Category 1 hurricane on October 29, 2020, reaching its peak intensity with winds of 170 knots on October 31, before dissipating on November 6, 2020. Our idealized simulation covered the period from October 28 to November 6, 2020. We initialized the model using ERA5 wind and wave data from 00:00 and 06:00 on October 28, along with wind data from 12:00. The idealized forecast was then driven by continuous ERA5 wind data input.

\begin{figure}
  \centering
  \noindent\includegraphics[width=\textwidth]{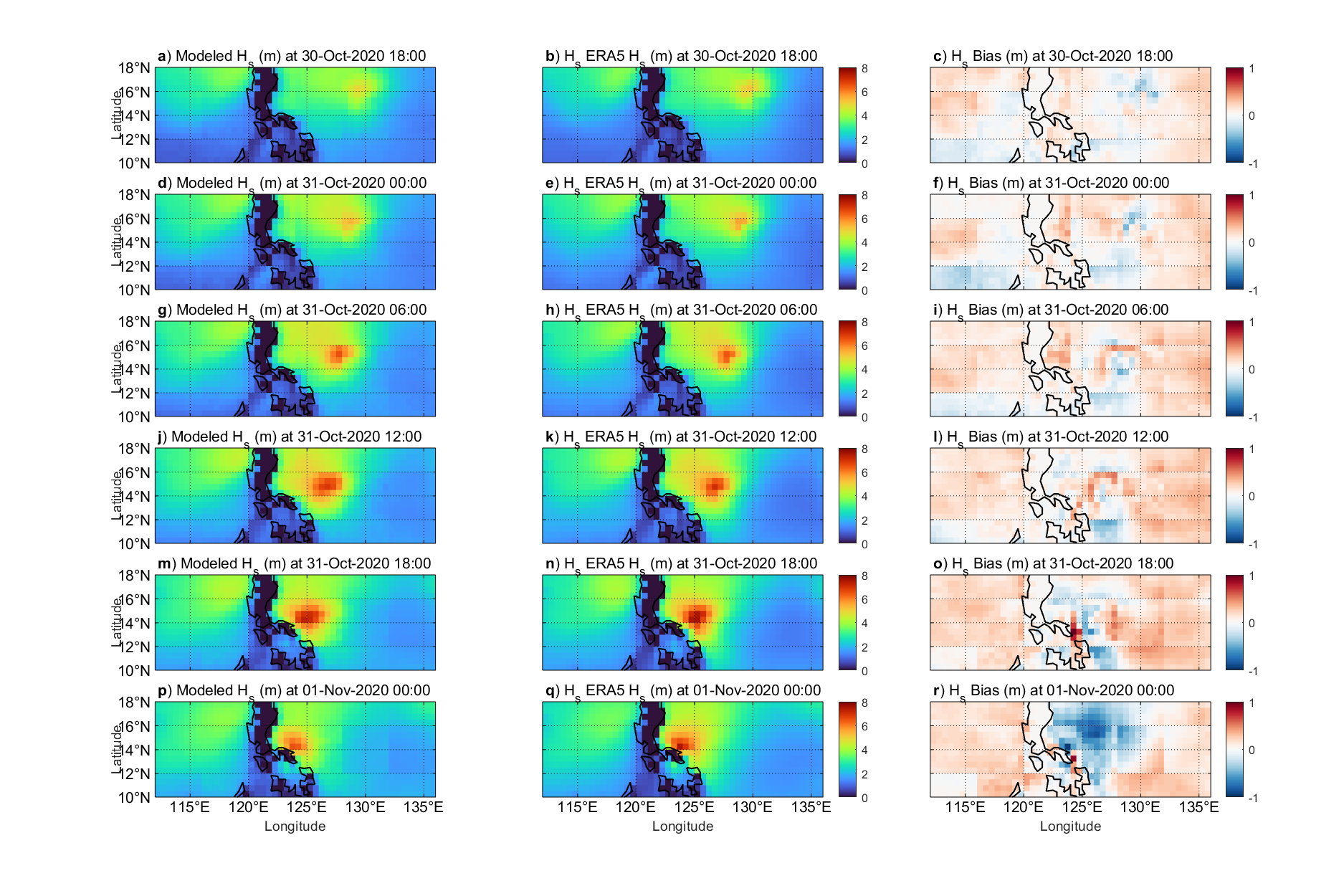}
  \caption{Model performance during the peak intensity phase of Typhoon Goni (October 30 - November 1, 2020). The figure presents a time series comparison of $H_s$ fields: (a,d,g,j,m,p) model predictions, (b,e,h,k,n,q) ERA5 reference data, and (c,f,i,l,o,r) bias maps (model minus ERA5) at six sequential time points. Bias maps use a diverging colormap where red indicates over-prediction and blue indicates under-prediction.}
  \label{fig:goni}
\end{figure}

Figure~\ref{fig:goni} provides a detailed visualization of the model's performance during Typhoon Goni's peak intensity phase. The figure presents a side-by-side comparison between our model predictions and ERA5 reference data for significant wave height at six time points spanning from October 30 at 18:00 to November 1 at 00:00, capturing the period of maximum storm intensity. This visual comparison is complemented by bias maps (model minus ERA5) using a diverging colormap to highlight regions of over-prediction (red) and under-prediction (blue).

The model is able to reproduce both the magnitude and spatial distribution of significant wave heights throughout this critical period. As shown in the left and middle columns, the predicted wave patterns closely match those observed in the ERA5 data, accurately capturing the intense wave activity associated with the typhoon at its peak strength. The wave height maxima, which exceed 6 m near the storm center, and their spatial extents are well represented, with the model successfully tracking the typhoon's movement across the Western Pacific.

The bias maps in the right column reveal that prediction errors remain minimal across most of the domain, generally within ±0.5 m even in the dynamically complex regions near the typhoon center. Some localized bias patterns are evident, with slight variations between over-prediction and under-prediction as the storm evolves. However, these biases are small relative to the overall wave heights and do not significantly impact the model's representation of the storm's wave field.

This case study of Typhoon Goni provides evidence of the model's performance under extreme conditions. To quantitatively validate this performance, an aggregate analysis was conducted on 16 major tropical cyclones from the 2020 season. The model achieved a mean regional RMSE of 0.215 m across all storms, demonstrating that the performance shown for Typhoon Goni is representative. The full analysis is detailed in the Supplementary Material (Section S2). The model's ability to accurately reproduce wave fields during this intense typhoon demonstrates its potential value for operational forecasting of ocean waves during severe weather events.

\subsubsection{Site-Specific Model Validation}
\label{sec:NDBCeval}
\begin{figure}
  \centering
  \noindent\includegraphics[width=\textwidth]{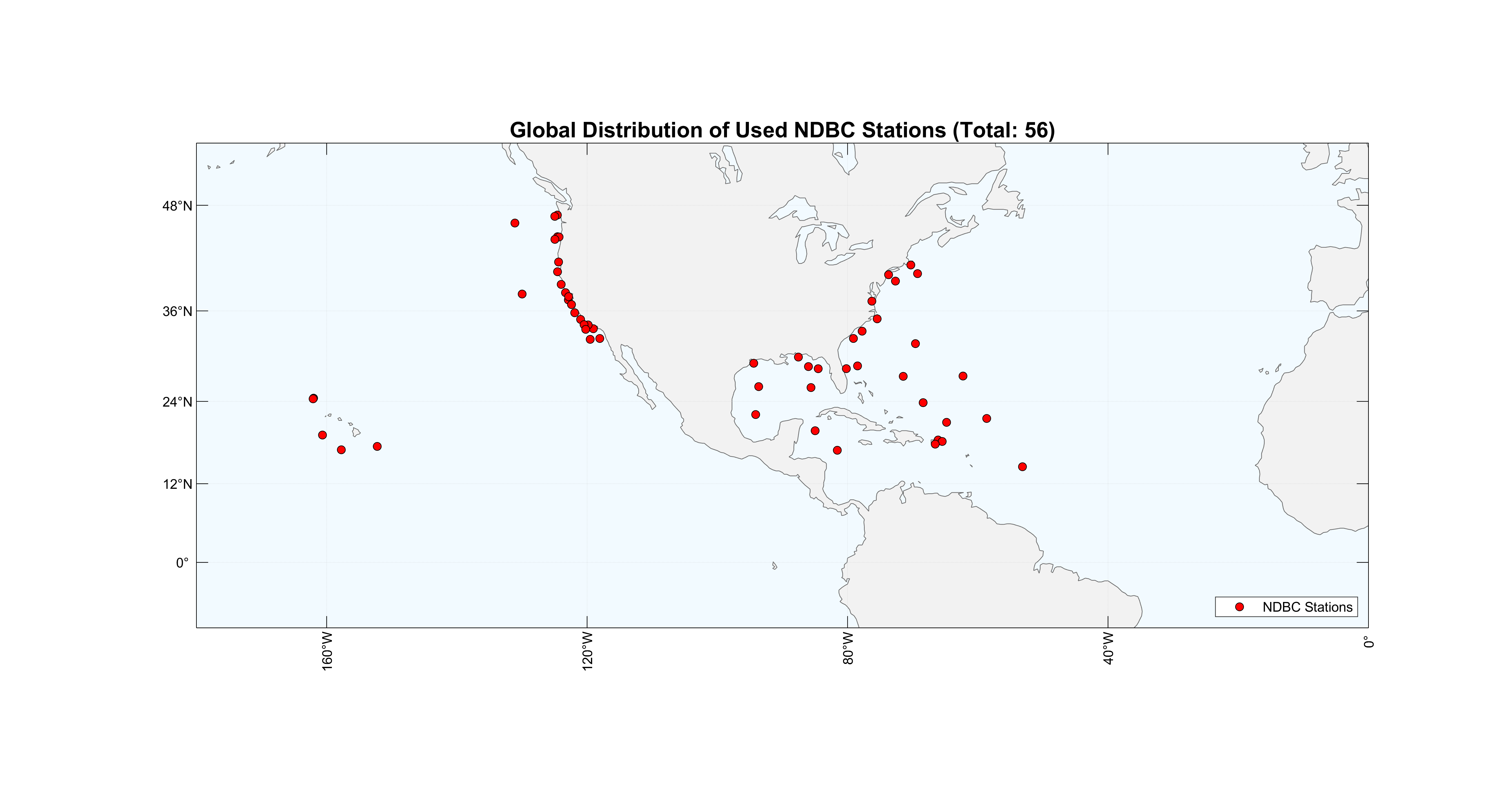}
  \caption{Global distribution of 56 National Data Buoy Center (NDBC) stations used for model validation in 2020. Red dots indicate the locations of the selected buoy stations.}
  \label{fig:site}
\end{figure}

To evaluate the model's performance against observational data, we utilized buoy data from the NDBC for the year 2020. We applied a filtration process to the data, excluding stations lacking significant wave height measurements and those with excessive data loss during the comparison period. This process resulted in a final set of 56 stations, as illustrated in Figure~\ref{fig:site}.

The model simulation followed the same protocol as the global evaluation. To ensure robust statistical analysis consistent with our global assessment, we conducted evaluations using 36 distinct forecast cases throughout the year. We initialized the model using ERA5 wind and wave data from 00:00 and 06:00 on the 1st, 4th, and 12th day of each month, along with wind data from 12:00, to predict wave conditions at 12:00. For each initialization, the simulation then continued for 59 additional steps using ERA5 wind data as input. NDBC data for the corresponding time periods were selected for comparison across all 36 forecast cases.

Two important considerations regarding the NDBC data must be noted. First, the period and direction measurements differ from our model outputs, limiting our comparison to significant wave height. Second, although the buoy data are reported 40 minutes past the hour (e.g., 10:40), these significant wave height values are considered representative of the conditions at the start of that hour (e.g., 10:00). This is because significant wave height is a statistical measure derived from a period of observation, typically the preceding hour. Therefore, we aligned these values with the corresponding hourly model outputs for our comparison, ensuring a temporally consistent analysis.

\begin{figure}
  \centering
  \noindent\includegraphics[width=\textwidth]{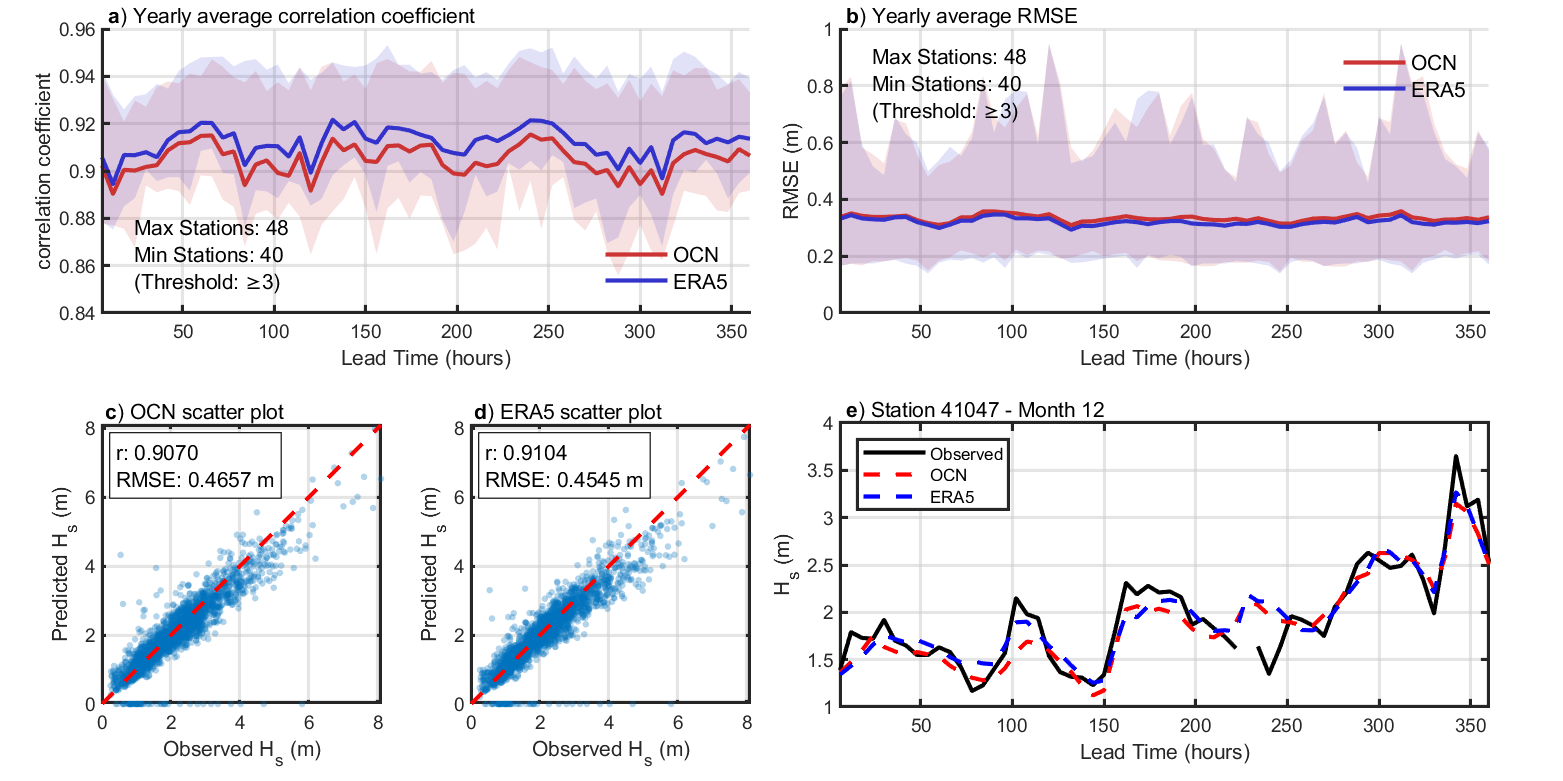}
  \caption{Model performance evaluation against NDBC buoy data for $H_s$ in 2020. (a) Yearly average correlation coefficient and (b) RMSE over 60 time steps across 36 forecast cases for both OCN (red) and ERA5 (blue). Solid lines represent mean values while shaded areas depict the interquartile range (25th to 75th percentiles). (c) Scatter plot of OCN predicted vs. observed $H_s$ and (d) ERA5 vs. observed $H_s$ for all stations and all forecast cases. (e) Time series comparison of observed (black), OCN (red), and ERA5 (blue) $H_s$ for Station 41047 for the forecast case initiated on December 12th, 2020.}
  \label{fig:position_metric}
\end{figure}

Figure~\ref{fig:position_metric} presents the comparison metrics for both OCN (red) and ERA5 (blue) against NDBC observations. Panels (a) and (b) illustrate the accuracy evolution over the forecast period for both datasets. Due to varying data availability across time steps, we indicate the minimum and maximum number of stations used in calculations for each step in the upper right corner. We computed monthly averages across all 36 forecast cases and implemented a threshold, excluding months with fewer than three available stations to mitigate the impact of extreme values. Solid lines represent mean values across all forecast cases, while shaded areas depict the interquartile range (25th to 75th percentiles), providing a more robust representation of forecast uncertainty compared to the full range of variability.

The results demonstrate that OCN achieves performance closely comparable to ERA5 throughout the prediction period. OCN maintains a mean correlation coefficient around 0.91 with RMSE around 0.35 m, while ERA5 shows correlation around 0.92 with RMSE around 0.32 m. This close agreement indicates that OCN successfully reproduces the wave height patterns present in ERA5 under idealized forecasting conditions. Panels (c) and (d) present scatter plots for OCN and ERA5 respectively, both showing close distributions along the ideal line with OCN achieving overall correlation of 0.91 (RMSE: 0.47 m) and ERA5 correlation of 0.91 (RMSE: 0.45 m).

Most importantly, panel (e) demonstrates OCN's ability to track real-time wave variations without exhibiting the lag often seen in static predictions, which can artificially inflate correlation coefficients. This specific case (Station 41047, December 2020) shows OCN closely following both observed wave dynamics and ERA5 patterns, confirming the model's capability to capture temporal wave evolution in dynamically rather than simply providing delayed responses.

\subsection{Comparative Evaluation with ECWAM Operational Forecasts}
In this section, we evaluate our model's performance under operational forecasting conditions by comparing it with ECMWF's ECWAM model, a state-of-the-art operational wave forecasting system. ECWAM is the wave component of ECMWF's Integrated Forecasting System (IFS), which provides global wave forecasts operationally and serves as a benchmark for wave prediction systems worldwide.

For this comparison, we conducted 36 forecast cases throughout 2024, initializing forecasts on the 1st, 4th, and 12th day of each month. While the operational ECWAM wave model produces forecasts four times daily (00:00, 06:00, 12:00, and 18:00 UTC), the corresponding ECMWF IFS wind forecast data available in the TIGGE archive is provided only for the 00:00 and 12:00 UTC cycles. To ensure that our model used the same atmospheric forcing, we restricted our analysis to the 10-day (240-hour) wave forecasts from the 00:00 UTC cycle. We resampled the ECWAM forecast data from its native variable resolution to a consistent 6-hour resolution for our analysis.

To initialize our model for a fair comparison, we used the first two time steps from the 00:00 UTC ECWAM operational forecast. Specifically, the ECWAM wave fields at forecast hour +0 (00:00 UTC) and +6 (06:00 UTC) were used as the historical wave state inputs for OCN. The model was then driven by the corresponding IFS wind forecasts from the same operational run to produce its own forecast, starting from forecast hour +12 (12:00 UTC). This initial prediction step leaves 38 remaining steps (228 hours) for a direct performance comparison between OCN and ECWAM.

To ensure a fair comparison between the models, all data was handled on a consistent 0.5° grid. In ECMWF's operational system, the ECWAM model is forced by high-resolution atmospheric data produced by the Integrated Forecasting System (IFS). For our OCN model, we used the corresponding IFS wind forecast data from the TIGGE archive, which is provided at a 0.5° resolution. The ECWAM operational wave data, which has a higher native resolution of 0.25°, was therefore resampled to our model's 0.5° grid for all comparative analyses.

\subsubsection{Performance Evaluation Using NDBC Observations}
\label{sec:NDBCcompare}
To assess model performance against in situ observations, we conducted a comprehensive station-by-station comparison using NDBC buoy data from 2024. Unlike the temporal evolution analysis presented earlier, this evaluation focuses on the relative performance of OCN and ECWAM at individual observation locations.

We employed a station-centric evaluation approach, extracting forecast data from both models at each NDBC station location and calculating performance metrics across all forecast lead times for each station. Only stations with sufficient data availability across all 12 months were retained for analysis to ensure robust statistical comparisons. For each qualifying station, we computed the overall correlation coefficient (r) and RMSE between model predictions and observations across the entire 228-hour forecast period. Stations were then classified based on their performance characteristics: those where OCN achieves both higher correlation coefficient and lower RMSE than ECWAM were considered OCN better, stations where ECWAM shows superior performance in both metrics were classified as ECWAM better, while remaining stations with mixed performance have one model excelling in correlation and the other in RMSE.

\begin{figure}
  \centering
  \noindent\includegraphics[width=\textwidth]{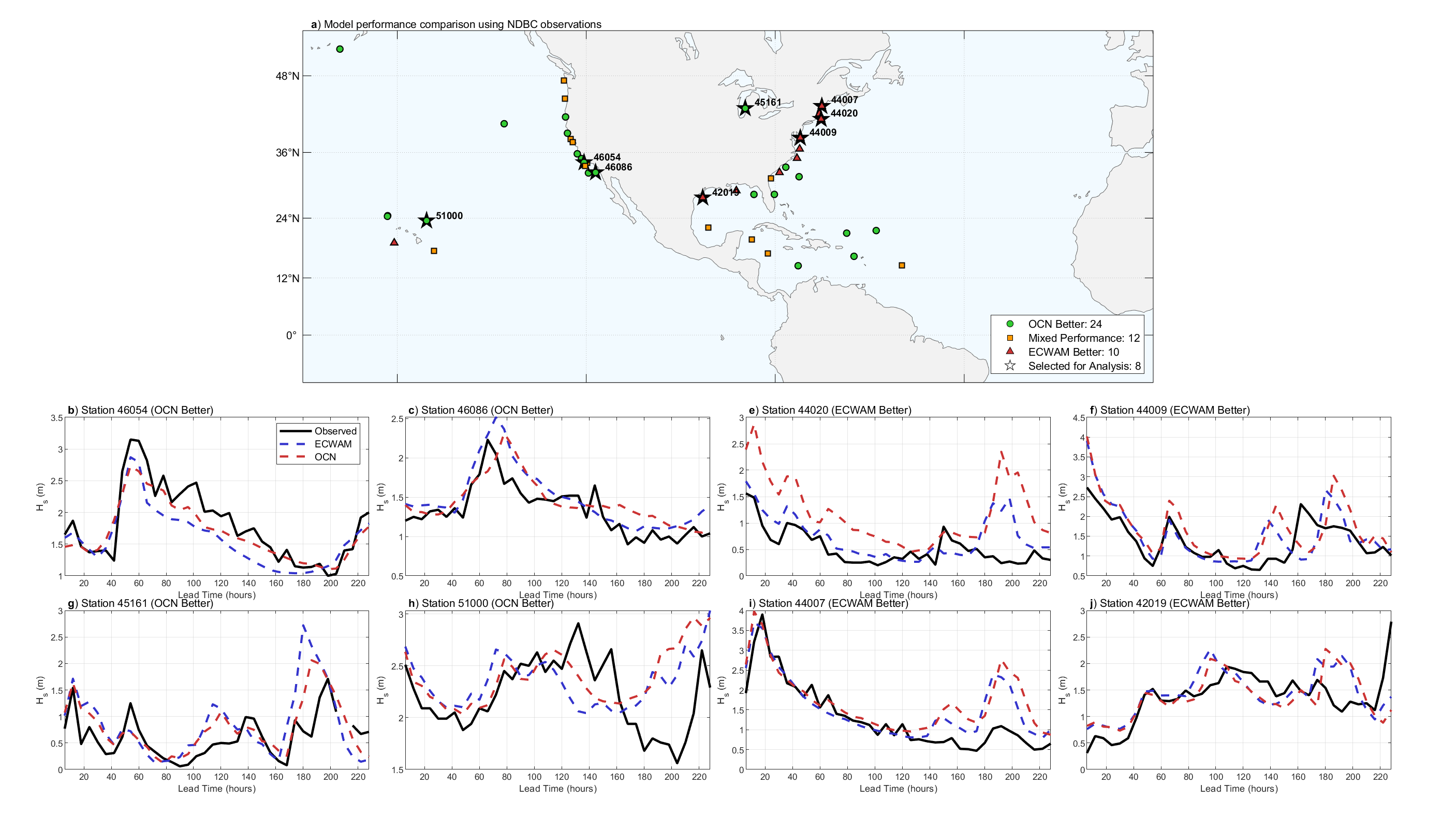}
  \caption{ Station-by-station performance comparison between OCN and ECWAM using NDBC observations. (a) Spatial distribution of performance categories showing stations where OCN performs better (green circles), ECWAM performs better (red triangles), mixed performance (orange squares), and selected stations for detailed analysis are marked with an unfilled star. (b-j) Time series comparisons for selected representative stations during the forecast case initiated April 12th, 2024, showing observed wave heights (black solid), ECWAM predictions (blue dashed), and OCN predictions (red dashed).}
  \label{fig:ndbc_comparison}
\end{figure}

Figure~\ref{fig:ndbc_comparison}, panel (a) presents the spatial distribution of station performance categories. The analysis reveals a relatively balanced performance landscape: OCN outperforms ECWAM at 24 stations, ECWAM shows superior performance at 10 stations, while 12 stations exhibit mixed performance characteristics. To examine performance differences in detail, we selected representative stations from each category using an integrated performance metric (r + 1/RMSE - 1). Panels (b-j) show time series comparisons for eight selected stations during a representative forecast case (initiated April 12th, 2024), with the left four panels showing stations where OCN performs better and the right four panels showing stations where ECWAM excels.

The time series reveal that both models generally capture the temporal evolution of wave height variations effectively. However, notable differences emerge at specific locations. For instance, at Station 44020 (panel e), OCN consistently predicts higher wave heights than both ECWAM and observations. This systematic bias can be attributed to limitations inherited from the ERA5 reanalysis data. Station 44020 is located in Nantucket Sound, a semi-enclosed water body that is naturally sheltered by Cape Cod to the north and the islands of Martha's Vineyard and Nantucket to the south. However, ERA5 reanalysis data also exhibits similar overestimation at this location compared to NDBC observations, indicating that OCN is faithfully reproducing the bias present in its training data. This difference can be attributed to the models' different spatial resolutions. The higher native resolution of ECWAM (0.25°) allows it to better resolve the complex coastal geometry and the wave-sheltering effects of the nearby islands, whereas OCN's lower resolution (0.5°) is inherited from the ERA5 training data, which also overestimates wave heights in this sheltered location.

The NDBC station-by-station analysis reveals that OCN performs comparably to ECWAM with a modest overall advantage. OCN achieves superior performance at 24 stations compared to ECWAM's 10 stations, with 12 stations showing mixed performance. While this indicates OCN's generally better accuracy at more locations, both models effectively capture temporal wave evolution across diverse coastal and oceanic environments. The performance differences suggest that OCN can achieve slightly better representation of observed wave conditions while maintaining comparable skill to the established physics-based ECWAM model.

\subsubsection{Performance Evaluation Using Jason-3 Observations}
\label{sec:Jasoncompare}
To provide an independent validation against satellite observations, we evaluated both models using Jason-3 altimeter data throughout 2024. Model performance was assessed through spatiotemporal collocation with Jason-3 satellite observations, which provide near-global coverage of significant wave height measurements with high spatial resolution (~7 km along-track spacing) and a 10-day repeat cycle.

For each model forecast time step, all Jason-3 observations falling within a ±0.5-hour temporal window were identified and spatially matched to the nearest model grid point. This collocation approach leverages the satellite's high spatial sampling rate to provide robust statistical validation across diverse oceanic conditions. Performance metrics were calculated using all collocated pairs for each forecast lead time, with a minimum threshold of 10 observations required for statistical significance. Quality control filters were applied to exclude land-contaminated measurements and remove unrealistic values (wave heights $<$ 0.1 m and exactly 0 m). The validation covers oceanic regions between 79.5°S and 80°N latitude, encompassing the overlapping coverage of both the satellite and model domains.

\begin{figure}
  \centering
  \noindent\includegraphics[width=\textwidth]{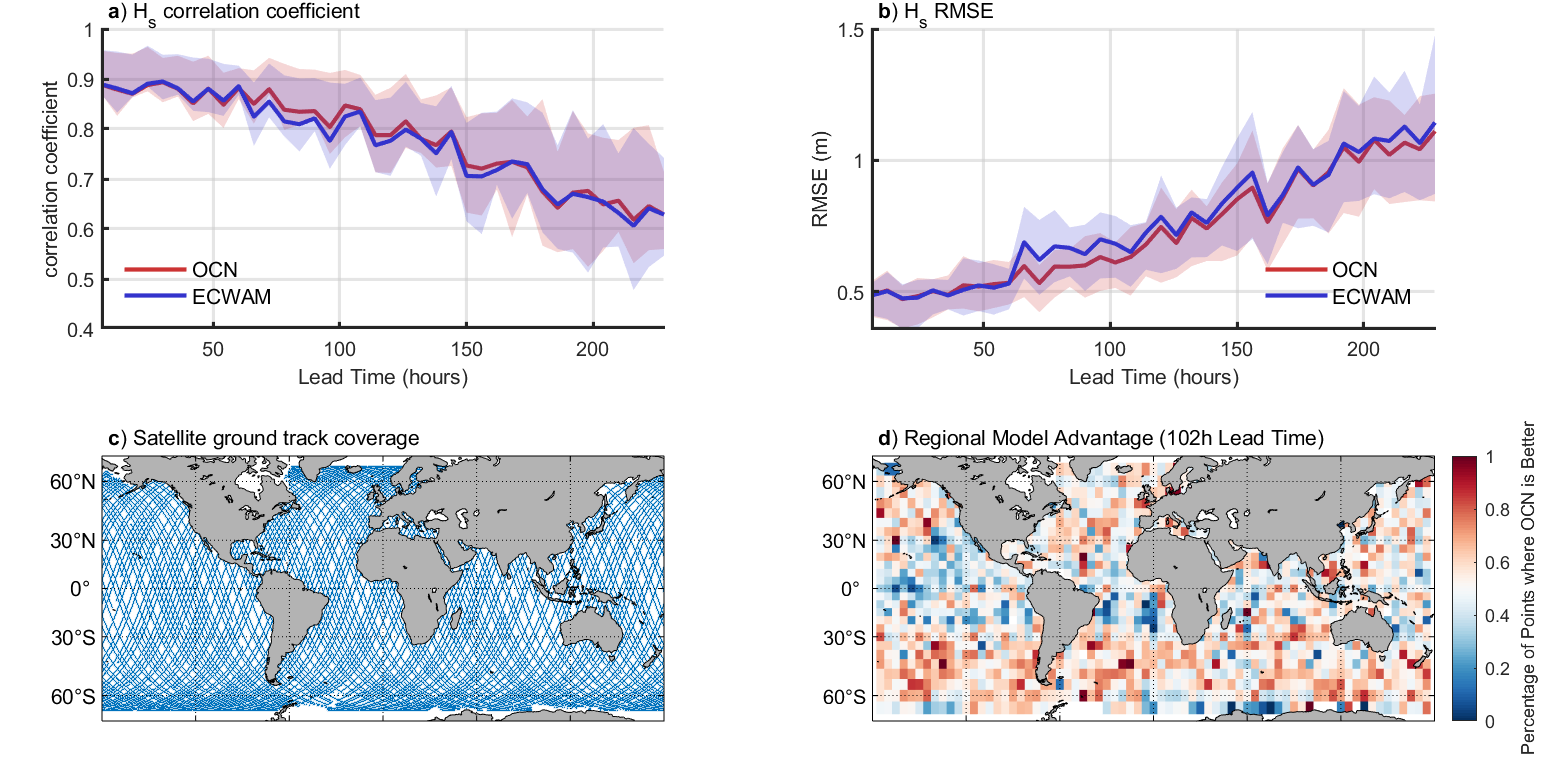}
  \caption{Comparison of OCN and ECWAM performance against Jason-3 satellite altimeter observations for 2024. (a) Correlation coefficient and (b) RMSE evolution over 228 hours of forecast lead time. (c) Global coverage of Jason-3 ground tracks used for validation (blue hatched areas). Solid lines represent mean values across all 36 forecast cases, with shaded areas indicating the interquartile range (25th to 75th percentiles). (d) Map of the Regional Model Advantage at a 102-hour lead time, showing the percentage of collocated points in each grid cell where OCN had a lower absolute error than ECWAM. Red indicates an advantage for OCN, while blue indicates an advantage for ECWAM.}
  \label{fig:jason3_comparison}
\end{figure}

Figure~\ref{fig:jason3_comparison} presents the comparative performance results. Panel (c) illustrates the global coverage of Jason-3 ground tracks used in the validation, demonstrating comprehensive sampling across all major ocean basins. The correlation coefficient evolution (panel a) shows that both models maintain similar performance throughout most of the forecast period, with OCN showing slight advantages in the 50-150 hour range. Both models achieve correlations above 0.8 at short lead times, declining to approximately 0.6-0.7 by 228 hours.

The RMSE comparison (panel b) reveals closely matched performance between the two models, with both exhibiting similar error growth patterns from approximately 0.5 m at initial forecast times to around 1.1 m at 228 hours. This near-identical RMSE evolution suggests that both models have comparable accuracy when validated against independent satellite observations.

To investigate the spatial distribution of model performance, we computed a regional model advantage metric at a 102-hour lead time (Figure~\ref{fig:jason3_comparison}, panel (d). The resulting map reveals complex, basin-scale patterns of performance. OCN shows a notable advantage in the West Pacific and in parts of the North Atlantic. Conversely, ECWAM demonstrates stronger performance in the East Pacific and the South Atlantic. In other major basins, such as the Indian and Southern Oceans, performance appears more mixed, with neither model showing a consistent, large-scale advantage. Overall, the performance between the two models is closely matched, with OCN holding a slight advantage in approximately 52\% of the analyzed oceanic regions. The detailed methodology for this spatial analysis is provided in the Supplementary Material (Section S3).

The Jason-3 satellite validation shows that OCN performs comparably to ECWAM with slight advantages in correlation metrics during the 50-150 hour forecast range, while maintaining equivalent RMSE throughout the prediction period. Both models exhibit similar error growth patterns and achieve comparable accuracy when validated against independent satellite observations. The results confirm that OCN can match the performance of operational wave forecasting systems while showing modest improvements in capturing large-scale oceanic wave patterns.

The comprehensive evaluation against independent observations demonstrates that OCN achieves performance comparable to the operational ECWAM system. Validation against NDBC buoys shows OCN performs better at more stations. The Jason-3 satellite comparison confirms this comparable skill but also reveals distinct regional strengths; OCN shows an advantage in the West Pacific, while ECWAM excels in the East Pacific and South Atlantic.

\section{Discussion}
Our results demonstrate that the OceanCastNet (OCN) model achieves forecasting performance comparable to the operational ECWAM system across multiple independent validation datasets. The analysis shows that a well-trained deep learning model can accurately reproduce complex wave dynamics, including during extreme weather events like tropical cyclones, while offering significant computational advantages.

It is important to note a subtle caveat in our operational forecast comparison. The ECMWF IFS atmospheric forcing used to drive our OCN model is part of a two-way coupled system with ECWAM. This means the wind fields are themselves influenced by the wave-induced surface roughness calculated by ECWAM. Consequently, our model indirectly benefits from the physical model it is being compared against. While running the IFS in an uncoupled mode was not feasible for this study, this inherent coupling is a relevant feature of the experimental design.

The computational cost of wave modeling has been a persistent challenge throughout the evolution of numerical wave prediction. Early wave models from the 1950s-1980s were severely limited by computational constraints, with first generation models unable to consider nonlinear wave interactions and second generation models only parameterizing these interactions. The development of third-generation spectral models like WAM (1984-1994) introduced explicit representation of wave physics without ad hoc spectral assumptions, but at substantial computational cost \cite{group1988wam}. Despite their physical sophistication, these models have traditionally been excluded from Earth system models due to prohibitive computational expenses \cite{ ikuyajolu2022porting}.

ECWAM, used as our primary comparison benchmark, and WAVEWATCH III represent the current state-of-the-art in operational wave forecasting, both being third-generation spectral models that solve similar governing equations with comparable computational architectures. WAVEWATCH III, developed at NOAA/NCEP as a further development of the WAM model, solves the random phase spectral action density balance equation for wavenumber-direction spectra \cite{tolmanUserManualSystem2009}. While ECWAM serves as ECMWF's operational wave forecasting component, specific computational cost metrics are not publicly available. However, WAVEWATCH III performance benchmarks provide representative computational requirements for this class of models: a 5-hour simulation using a global ocean with an unstructured mesh with 59,072 nodes (ranges from 1° at the Equator to 0.5° along coastlines) requires 6,689 seconds (1.86 hours) on a single IBM POWER9 CPU core \cite{ ikuyajolu2022porting}. This resolution is comparable to our model's 0.5° global resolution, making this benchmark highly relevant. Extrapolating these results, a 10-day global forecast would demand approximately 37 hours of single-core computation time.

Our deep learning approach achieves computational efficiency gains exceeding four orders of magnitude compared to traditional spectral models. OCN completes equivalent 10-day global wave forecasts in approximately 10 seconds on consumer-grade GPU hardware (RTX 3060 mobile), compared to the 37+ hours required by CPU-based spectral approaches. This dramatic improvement transcends conventional optimization strategies, representing a fundamental algorithmic advancement that maintains physical consistency through explicit external forcing integration. While traditional GPU acceleration of spectral models yields modest 2-4× speedups \cite{ ikuyajolu2022porting}, our methodology enables deployment on readily available hardware, democratizing access to high-quality wave forecasting capabilities previously restricted to well-resourced computational centers.

The computational efficiency transformation—from hours of supercomputer time to seconds on consumer hardware—fundamentally expands the scope of feasible wave modeling applications. This advancement enables real-time ensemble forecasting, uncertainty quantification, and operational deployment scenarios previously constrained by resource limitations. Our GPU-native approach positions wave forecasting to fully exploit modern computing capabilities while maintaining accessibility for broader scientific community participation.

\section{Conclusion}
This study presents OceanCastNet (OCN), a deep learning alternative to conventional wave forecasting models such as ECWAM. OCN demonstrates that machine learning approaches can achieve comparable accuracy to established physics-based wave models while offering significant computational advantages.

Comprehensive evaluation against multiple datasets shows OCN performs competitively with the operational ECWAM system. Comparisons using ERA5 reanalysis data reveal OCN maintains consistently higher anomaly correlation coefficients for significant wave height, particularly after 100 hours, with similar RMSE values. Station-by-station validation using NDBC buoy observations shows OCN outperforming ECWAM at 24 stations compared to 10 stations where ECWAM excels. Independent validation using Jason-3 satellite data confirms comparable performance between both models, while also revealing distinct regional strengths, with OCN showing an advantage in the West Pacific and ECWAM excelling in other basins. The model successfully handles extreme weather conditions, as demonstrated during Typhoon Goni with prediction errors generally within ±0.5 m.

OCN provides a computationally efficient alternative to conventional wave models, completing forecasts in seconds compared to hours required by traditional spectral approaches. This efficiency enables potential applications in real-time forecasting scenarios where computational resources are limited.

The results demonstrate that machine learning wave prediction can serve as a viable alternative to conventional wave forecasting systems, offering comparable accuracy with substantial computational benefits for operational wave prediction applications.

\appendix
\section{Appendix: Evaluation Metrics and Model Hyperparameters}
\label{app:Formulas}
We employed the latitude-weighted anomaly correlation coefficient (ACC), root mean square error (RMSE), and mean error (ME) to evaluate the model performance when the target data was ERA5. The definitions of these metrics are derived from \cite{raspWeatherBenchBenchmarkData2020a}, but we adopted the slightly modified versions from \cite{pathakFourcastnetGlobalDatadriven2022}. The latitude-weighted ACC is defined as follows:
\begin{equation}
ACC(v,l)=\frac{\sum_{m,n} L(m) \tilde{X}{pred}{(l)}[v,m,n] \tilde{X}{true}{(l)}[v,m,n]}{\sqrt{\sum_{m,n} L(m) (\tilde{X}{pred}{(l)}[v,m,n])^2 \sum_{m,n} L(m) (\tilde{X}_{true}{(l)}[v,m,n])^2}},
\end{equation}
where $\tilde{X}{pred/true}{(l)}$ represents the long-term-mean-subtracted value of predicted (/true) variable $v$ at the location denoted by the grid coordinates $(m,n)$ at the forecast time step $l$. The long-term mean of a variable is simply the mean value of that variable over a large number of historical samples in the training dataset. The long-term mean-subtracted variables $\tilde{X}{pred/true}$ represent the anomalies of those variables that are not captured by the long-term mean values. $L(m)$ is the latitude weighting factor at the coordinate $m$, defined as follows:
\begin{equation}
L(j)=\frac{\cos(lat(m))}{\frac{1}{N_{lat}} \sum_j^{N_{lat}} \cos(lat(m))}.
\end{equation}
The latitude-weighted RMSE for a forecast variable $v$ at forecast time step $l$ is defined by the following equation, using the same latitude weighting factor as ACC:
\begin{equation}
RMSE(v,l)=\sqrt{\frac{1}{NM} \sum_{m=1}^M \sum_{n=1}^N L(m) (X_{pred}{(l)}[v,m,n]-X_{true}{(l)}[v,m,n])^2},
\end{equation}
where $X_{pred/true}^{(l)}[v,m,n]$ represents the value of predicted (/true) variable $v$ at the location denoted by the grid coordinates $(m,n)$ at the forecast time step $l$.
The mean error (ME) for a forecast variable $v$ at forecast time step $l$ is defined by the following equation: \begin{equation} ME(v,l)[m,n] = \frac{1}{K} \sum_{k=1}^{K} (X_{pred,k}^{(l)}[v,m,n] - X_{true,k}^{(l)}[v,m,n]), \end{equation} where $X_{pred/true,k}^{(l)}[v,m,n]$ represents the value of predicted (/true) variable $v$ at the location denoted by the grid coordinates $(m,n)$ at the forecast time step $l$ for forecast case $k$, and $K$ is the total number of forecast cases. By preserving the sign of the differences, ME reveals spatial patterns of systematic bias, indicating regions where the model tends to overestimate (positive values) or underestimate (negative values) the forecasted variable.

When evaluating model performance against observational data (NDBC buoy data and Jason-3 satellite observations), we employed two different approaches for calculating the Pearson correlation coefficient (r) and root mean square error (RMSE) depending on the analysis objective.

Time-step approach (used in sections 3.1.2 and 3.2.2): For each time step, we calculate r and RMSE across all available sites or observations. This approach is used when evaluating temporal evolution of model performance. The Pearson correlation coefficient measures the linear correlation between predicted and observed values, ranging from -1 to 1, where 1 indicates perfect positive correlation, 0 indicates no linear correlation, and -1 indicates perfect negative correlation. The RMSE quantifies the average magnitude of prediction errors. For each time step t, these metrics are calculated as follows:

\begin{equation}
r(t) = \frac{\sum_{i=1}^{N} (X_{pred,i}^{(t)} - \bar{X}_{pred}^{(t)})(X_{obs,i}^{(t)} - \bar{X}_{obs}^{(t)})}{\sqrt{\sum_{i=1}^{N} (X_{pred,i}^{(t)} - \bar{X}_{pred}^{(t)})^2 \sum_{i=1}^{N} (X_{obs,i}^{(t)} - \bar{X}_{obs}^{(t)})^2}}
\end{equation}
\begin{equation}
RMSE(t) = \sqrt{\frac{1}{N} \sum_{i=1}^{N} (X_{pred,i}^{(t)} - X_{obs,i}^{(t)})^2}
\end{equation}

where $X_{pred,i}^{(t)}$ and $X_{obs,i}^{(t)}$ represent the predicted and observed values at site i and time step t, respectively. $\bar{X}_{pred}^{(t)}$ and $\bar{X}_{obs}^{(t)}$ are the means of predicted and observed values across all sites at time step t, and N is the total number of available sites at that time step.

Station-centric approach (used in section 3.2.1): For each station, we calculate r and RMSE across all available time steps. This approach is used when comparing model performance at individual locations. For each station s, these metrics are calculated as follows:

\begin{equation}
r(s) = \frac{\sum_{t=1}^{T} (X_{pred,s}^{(t)} - \bar{X}_{pred,s})(X_{obs,s}^{(t)} - \bar{X}_{obs,s})}{\sqrt{\sum_{t=1}^{T} (X_{pred,s}^{(t)} - \bar{X}_{pred,s})^2 \sum_{t=1}^{T} (X_{obs,s}^{(t)} - \bar{X}_{obs,s})^2}}
\end{equation}
\begin{equation}
RMSE(s) = \sqrt{\frac{1}{T} \sum_{t=1}^{T} (X_{pred,s}^{(t)} - X_{obs,s}^{(t)})^2}
\end{equation}

where $X_{pred,s}^{(t)}$ and $X_{obs,s}^{(t)}$ represent the predicted and observed values at station s and time step t, respectively. $\bar{X}_{pred,s}$ and $\bar{X}_{obs,s}$ are the means of predicted and observed values across all time steps at station s, and T is the total number of available time steps at that station.

The full architectural and training hyperparameter settings for the OceanCastNet model, as referenced in Section 2.2 , are detailed in Table ~\ref{tab:hyperparameters}.

\begin{table}[h!]
\caption{Hyperparameter settings for the OceanCastNet model.}
\label{tab:hyperparameters}
\centering
\begin{tabular}{l c}
\hline
\textbf{Parameter} & \textbf{Value / Setting} \\
\hline
\multicolumn{2}{l}{\textit{Architectural Hyperparameters}} \\
Image Size & (320, 720) \\
Patch Size & (8, 8) \\
Input Channels & 12 \\
Output Channels & 3 \\
Embedding Dimension & 768 \\
Depth (AFNO Layers) & 12 \\
MLP Ratio & 4.0 \\
Spatial Mixing Blocks & 8 \\
\hline
\multicolumn{2}{l}{\textit{Training Hyperparameters}} \\
Learning Rate & 5e-4 \\
Batch Size & 32 \\
Max Epochs & 50 \\
LR Scheduler & CosineAnnealingLR \\
\hline
\multicolumn{2}{l}{\textit{Regularization Hyperparameters}} \\
Drop Rate & 0.0 \\
Drop Path Rate & 0.0 \\
Sparsity Threshold & 0.01 \\
Hard Thresholding Fraction & 1.0 \\
\hline
\end{tabular}
\end{table}

%%%%%%%%%%%%%%%%%%%%%%%%%%%%%%%%%%%%%%%%%%%%%%%
% Optional Glossary, Notation or Acronym section goes here:
%
% Glossary is only allowed in Reviews of Geophysics
%  \begin{glossary}
%  \term{Term}
%   Term Definition here
%  \term{Term}
%   Term Definition here
%  \term{Term}
%   Term Definition here
%  \end{glossary}

%%%%%%%%%%%%%%%%%%%%%%%%%%%%%%%%%%%%%%%%%%%%%%%
% Acronyms
%% NOTE that acronyms in the final published version will be spelled out when used in figure captions.
%   \begin{acronyms}
%   \acro{Acronym}
%   Definition here
%   \acro{EMOS}
%   Ensemble model output statistics
%   \acro{ECMWF}
%   Centre for Medium-Range Weather Forecasts
%   \end{acronyms}

%%%%%%%%%%%%%%%%%%%%%%%%%%%%%%%%%%%%%%%%%%%%%%%
% Notation
%   \begin{notation}
%   \notation{$a+b$} Notation Definition here
%   \notation{$e=mc^2$}
%   Equation in German-born physicist Albert Einstein's theory of special
%  relativity that showed that the increased relativistic mass ($m$) of a
%  body comes from the energy of motion of the body—that is, its kinetic
%  energy ($E$)—divided by the speed of light squared ($c^2$).
%   \end{notation}

%%%%%%%%%%%%%%%%%%%%%%%%%%%%%%%%%%%%%%%%%%%%%%%
%
% DATA SECTION and ACKNOWLEDGMENTS
%
%%%%%%%%%%%%%%%%%%%%%%%%%%%%%%%%%%%%%%%%%%%%%%%

\section*{Open Research Section}
The ERA5 reanalysis data, including 10-meter wind speed, significant wave height, mean wave period and mean wave direction, were obtained from the Copernicus Climate Change Service (C3S) Climate Data Store \cite{hersbachERA5HourlyData2023}. The IFS data used for comparison were acquired from the THORPEX Interactive Grand Global Ensemble (TIGGE) dataset \cite{tiggeDataset2010}. As of June 2024, ECMWF is in the process of migrating this dataset to a new service, and the original link may be inactive. Jason-3 altimeter data were obtained from NASA's Physical Oceanography Distributed Active Archive Center (PO.DAAC) \cite{desaiJason3GPSBased2016}. Additional observational data were acquired from the National Data Buoy Center \cite{ndbcData2024}. The evaluation datasets, plotting code for figure generation, model source code, trained parameters, and a running example for this study are openly available in Zenodo \cite{Zhang2025zenodo}.

\section*{Conflict of Interest declaration}
The authors declare there are no conflicts of interest for this manuscript.

\acknowledgments

We would like to express our gratitude to the European Centre for Medium‐Range Weather Forecasts (ECMWF) for providing the ERA5 reanalysis data used for model training. We would like to thank the Dawning Information Industry Co., Ltd. for providing the computational resources used in this study. We extend our special thanks to the authors of FourCastNet for their inspiring work. The architecture of ou2r OceanCastNet model is built upon the framework of FourCastNet, and we are grateful for their innovative contributions to the field of deep learning‐based weather forecasting.

This work was supported by the Key R\&D Program of Shandong Province, China (grant number 2024SFGC0201), the Dongtou District Science and Technology R\&D project (grant number S2023Y09), the National Key Research and Development Program of China (grant number 2022YFD2401304), the Guanghe fund (grant number ghfund202407033074), and the National Social Science Fund of China (grant number 23CGL008) (to H.Y.); and by the Qingdao Science and Technology Innovation Strategy Research Program Project (grant number 25-1-4-zlyj-7-zhc) and the Shandong Province Higher Education Institutions Youth Innovation and Technology Support Program (grant number 2024KJN017) (to X.Q.).

We thank the editor and two anonymous reviewers for their constructive feedback, which significantly improved this manuscript.
%%%%%%%%%%%%%%%%%%%%%%%%%%%%%%%%%%%%%%%%%%%%%%%
% REFERENCES and BIBLIOGRAPHY
%
% \bibliography{<name of your .bib file>} don't specify the file extension
% don't specify bibliographystyle
%
\bibliography{cite.bib}

\begin{thebibliography}{}

\bibitem [\protect \citeauthoryear {%
Adytia%
, Saepudin%
, Pudjaprasetya%
, Husrin%
\BCBL {}\ \BBA {} Sopaheluwakan%
}{%
Adytia%
\ \protect \BOthers {.}}{%
{\protect \APACyear {2022}}%
}]{%
adytiaDeepLearningApproach2022}
\APACinsertmetastar {%
adytiaDeepLearningApproach2022}%
\begin{APACrefauthors}%
Adytia, D.%
, Saepudin, D.%
, Pudjaprasetya, S\BPBI R.%
, Husrin, S.%
\BCBL {}\ \BBA {} Sopaheluwakan, A.%
\end{APACrefauthors}%
\unskip\
\newblock
\APACrefYearMonthDay{2022}{{\APACmonth{01}}}{}.
\newblock
{\BBOQ}\APACrefatitle {A {{Deep Learning Approach}} for {{Wave Forecasting
  Based}} on a {{Spatially Correlated Wind Feature}}, with a {{Case Study}} in
  the {{Java Sea}}, {{Indonesia}}} {A {{Deep Learning Approach}} for {{Wave
  Forecasting Based}} on a {{Spatially Correlated Wind Feature}}, with a {{Case
  Study}} in the {{Java Sea}}, {{Indonesia}}}.{\BBCQ}
\newblock
\APACjournalVolNumPages{Fluids}{7}{1}{39}.
\newblock
\begin{APACrefDOI} \doi{10.3390/fluids7010039} \end{APACrefDOI}
\PrintBackRefs{\CurrentBib}

\bibitem [\protect \citeauthoryear {%
Bai%
, Wang%
, Zhu%
\BCBL {}\ \BBA {} Feng%
}{%
Bai%
\ \protect \BOthers {.}}{%
{\protect \APACyear {2022}}%
}]{%
bai2022development}
\APACinsertmetastar {%
bai2022development}%
\begin{APACrefauthors}%
Bai, G.%
, Wang, Z.%
, Zhu, X.%
\BCBL {}\ \BBA {} Feng, Y.%
\end{APACrefauthors}%
\unskip\
\newblock
\APACrefYearMonthDay{2022}{}{}.
\newblock
{\BBOQ}\APACrefatitle {Development of a 2-{{D}} Deep Learning Regional Wave
  Field Forecast Model Based on Convolutional Neural Network and the
  Application in {{South China Sea}}} {Development of a 2-{{D}} deep learning
  regional wave field forecast model based on convolutional neural network and
  the application in {{South China Sea}}}.{\BBCQ}
\newblock
\APACjournalVolNumPages{Applied Ocean Research}{118}{}{103012}.
\PrintBackRefs{\CurrentBib}

\bibitem [\protect \citeauthoryear {%
Bi%
\ \protect \BOthers {.}}{%
Bi%
\ \protect \BOthers {.}}{%
{\protect \APACyear {2022}}%
}]{%
biPanguweather3dHighresolution2022}
\APACinsertmetastar {%
biPanguweather3dHighresolution2022}%
\begin{APACrefauthors}%
Bi, K.%
, Xie, L.%
, Zhang, H.%
, Chen, X.%
, Gu, X.%
\BCBL {}\ \BBA {} Tian, Q.%
\end{APACrefauthors}%
\unskip\
\newblock
\APACrefYearMonthDay{2022}{}{}.
\newblock
{\BBOQ}\APACrefatitle {Pangu-Weather: {{A}} 3d High-Resolution Model for Fast
  and Accurate Global Weather Forecast} {Pangu-weather: {{A}} 3d
  high-resolution model for fast and accurate global weather forecast}.{\BBCQ}
\newblock
\APACjournalVolNumPages{arXiv preprint arXiv:2211.02556}{}{}{}.
\PrintBackRefs{\CurrentBib}

\bibitem [\protect \citeauthoryear {%
Desai%
}{%
Desai%
}{%
{\protect \APACyear {2016}}%
}]{%
desaiJason3GPSBased2016}
\APACinsertmetastar {%
desaiJason3GPSBased2016}%
\begin{APACrefauthors}%
Desai, S.%
\end{APACrefauthors}%
\unskip\
\newblock
\APACrefYearMonthDay{2016}{}{}.
\newblock
\APACrefbtitle {{Jason-3 GPS based orbit and SSHA OGDR [Dataset]}.} {{Jason-3
  GPS based orbit and SSHA OGDR [Dataset]}.}
\newblock
\APACaddressPublisher{}{{NASA Physical Oceanography Distributed Active Archive
  Center}}.
\newblock
\begin{APACrefURL}
  \url{https://podaac.jpl.nasa.gov/dataset/JASON_3_L2_OST_OGDR_GPS}
  \end{APACrefURL}
\newblock
\begin{APACrefDOI} \doi{10.5067/J3L2G-OGDRF} \end{APACrefDOI}
\PrintBackRefs{\CurrentBib}

\bibitem [\protect \citeauthoryear {%
Group%
}{%
Group%
}{%
{\protect \APACyear {1988}}%
}]{%
group1988wam}
\APACinsertmetastar {%
group1988wam}%
\begin{APACrefauthors}%
Group, T\BPBI W.%
\end{APACrefauthors}%
\unskip\
\newblock
\APACrefYearMonthDay{1988}{}{}.
\newblock
{\BBOQ}\APACrefatitle {The WAM model—A third generation ocean wave prediction
  model} {The wam model—a third generation ocean wave prediction
  model}.{\BBCQ}
\newblock
\APACjournalVolNumPages{Journal of physical oceanography}{18}{12}{1775--1810}.
\PrintBackRefs{\CurrentBib}

\bibitem [\protect \citeauthoryear {%
Guibas%
\ \protect \BOthers {.}}{%
Guibas%
\ \protect \BOthers {.}}{%
{\protect \APACyear {2021}}%
}]{%
guibasAdaptiveFourierNeural2021}
\APACinsertmetastar {%
guibasAdaptiveFourierNeural2021}%
\begin{APACrefauthors}%
Guibas, J.%
, Mardani, M.%
, Li, Z.%
, Tao, A.%
, Anandkumar, A.%
\BCBL {}\ \BBA {} Catanzaro, B.%
\end{APACrefauthors}%
\unskip\
\newblock
\APACrefYearMonthDay{2021}{}{}.
\newblock
{\BBOQ}\APACrefatitle {Adaptive Fourier Neural Operators: {{Efficient}} Token
  Mixers for Transformers} {Adaptive fourier neural operators: {{Efficient}}
  token mixers for transformers}.{\BBCQ}
\newblock
\APACjournalVolNumPages{arXiv preprint arXiv:2111.13587}{}{}{}.
\PrintBackRefs{\CurrentBib}

\bibitem [\protect \citeauthoryear {%
Hersbach%
\ \protect \BOthers {.}}{%
Hersbach%
\ \protect \BOthers {.}}{%
{\protect \APACyear {2023}}%
}]{%
hersbachERA5HourlyData2023}
\APACinsertmetastar {%
hersbachERA5HourlyData2023}%
\begin{APACrefauthors}%
Hersbach, H.%
, Bell, B.%
, Berrisford, P.%
, Biavati, G.%
, Horanyi, A.%
, Munoz~Sabater, J.%
\BDBL {}Thepaut, J\BHBI N.%
\end{APACrefauthors}%
\unskip\
\newblock
\APACrefYearMonthDay{2023}{}{}.
\newblock
\APACrefbtitle {{ERA5 hourly data on single levels from 1940 to present
  [Dataset]}.} {{ERA5 hourly data on single levels from 1940 to present
  [Dataset]}.}
\newblock
\APACaddressPublisher{}{{Copernicus Climate Change Service (C3S) Climate Data
  Store (CDS)}}.
\newblock
\begin{APACrefURL}
  \url{https://cds.climate.copernicus.eu/datasets/reanalysis-era5-single-levels}
  \end{APACrefURL}
\newblock
\begin{APACrefDOI} \doi{10.24381/cds.adbb2d47} \end{APACrefDOI}
\PrintBackRefs{\CurrentBib}

\bibitem [\protect \citeauthoryear {%
Hersbach%
\ \protect \BOthers {.}}{%
Hersbach%
\ \protect \BOthers {.}}{%
{\protect \APACyear {2020}}%
}]{%
hersbachERA5GlobalReanalysis2020}
\APACinsertmetastar {%
hersbachERA5GlobalReanalysis2020}%
\begin{APACrefauthors}%
Hersbach, H.%
, Bell, B.%
, Berrisford, P.%
, Hirahara, S.%
, Hor{\'a}nyi, A.%
, Mu{\~n}oz-Sabater, J.%
\BDBL {}Schepers, D.%
\end{APACrefauthors}%
\unskip\
\newblock
\APACrefYearMonthDay{2020}{}{}.
\newblock
{\BBOQ}\APACrefatitle {The {{ERA5}} Global Reanalysis} {The {{ERA5}} global
  reanalysis}.{\BBCQ}
\newblock
\APACjournalVolNumPages{Quarterly Journal of the Royal Meteorological
  Society}{146}{730}{1999--2049}.
\PrintBackRefs{\CurrentBib}

\bibitem [\protect \citeauthoryear {%
Ikuyajolu%
, Van~Roekel%
, Brus%
, Thomas%
\BCBL {}\ \BBA {} Deng%
}{%
Ikuyajolu%
\ \protect \BOthers {.}}{%
{\protect \APACyear {2022}}%
}]{%
ikuyajolu2022porting}
\APACinsertmetastar {%
ikuyajolu2022porting}%
\begin{APACrefauthors}%
Ikuyajolu, O\BPBI J.%
, Van~Roekel, L.%
, Brus, S\BPBI R.%
, Thomas, E\BPBI E.%
\BCBL {}\ \BBA {} Deng, Y.%
\end{APACrefauthors}%
\unskip\
\newblock
\APACrefYearMonthDay{2022}{}{}.
\newblock
{\BBOQ}\APACrefatitle {Porting the WAVEWATCH III (v6. 07) wave action source
  terms to GPU} {Porting the wavewatch iii (v6. 07) wave action source terms to
  gpu}.{\BBCQ}
\newblock
\APACjournalVolNumPages{Geoscientific Model Development
  Discussions}{2022}{}{1--26}.
\PrintBackRefs{\CurrentBib}

\bibitem [\protect \citeauthoryear {%
Jing%
, Zhang%
, Hao%
\BCBL {}\ \BBA {} Huang%
}{%
Jing%
\ \protect \BOthers {.}}{%
{\protect \APACyear {2022}}%
}]{%
jing2022numerical}
\APACinsertmetastar {%
jing2022numerical}%
\begin{APACrefauthors}%
Jing, Y.%
, Zhang, L.%
, Hao, W.%
\BCBL {}\ \BBA {} Huang, L.%
\end{APACrefauthors}%
\unskip\
\newblock
\APACrefYearMonthDay{2022}{}{}.
\newblock
{\BBOQ}\APACrefatitle {Numerical Study of a {{CNN-based}} Model for Regional
  Wave Prediction} {Numerical study of a {{CNN-based}} model for regional wave
  prediction}.{\BBCQ}
\newblock
\APACjournalVolNumPages{Ocean Engineering}{255}{}{111400}.
\PrintBackRefs{\CurrentBib}

\bibitem [\protect \citeauthoryear {%
Lam%
\ \protect \BOthers {.}}{%
Lam%
\ \protect \BOthers {.}}{%
{\protect \APACyear {2022}}%
}]{%
lamGraphCastLearningSkillful2022}
\APACinsertmetastar {%
lamGraphCastLearningSkillful2022}%
\begin{APACrefauthors}%
Lam, R.%
, {Sanchez-Gonzalez}, A.%
, Willson, M.%
, Wirnsberger, P.%
, Fortunato, M.%
, Pritzel, A.%
\BDBL {}others%
\end{APACrefauthors}%
\unskip\
\newblock
\APACrefYearMonthDay{2022}{}{}.
\newblock
{\BBOQ}\APACrefatitle {{{GraphCast}}: {{Learning}} Skillful Medium-Range Global
  Weather Forecasting} {{{GraphCast}}: {{Learning}} skillful medium-range
  global weather forecasting}.{\BBCQ}
\newblock
\APACjournalVolNumPages{arXiv preprint arXiv:2212.12794}{}{}{}.
\PrintBackRefs{\CurrentBib}

\bibitem [\protect \citeauthoryear {%
Li%
\ \protect \BOthers {.}}{%
Li%
\ \protect \BOthers {.}}{%
{\protect \APACyear {2020}}%
}]{%
liFourierNeuralOperator2020}
\APACinsertmetastar {%
liFourierNeuralOperator2020}%
\begin{APACrefauthors}%
Li, Z.%
, Kovachki, N.%
, Azizzadenesheli, K.%
, Liu, B.%
, Bhattacharya, K.%
, Stuart, A.%
\BCBL {}\ \BBA {} Anandkumar, A.%
\end{APACrefauthors}%
\unskip\
\newblock
\APACrefYearMonthDay{2020}{}{}.
\newblock
{\BBOQ}\APACrefatitle {Fourier Neural Operator for Parametric Partial
  Differential Equations} {Fourier neural operator for parametric partial
  differential equations}.{\BBCQ}
\newblock
\APACjournalVolNumPages{arXiv preprint arXiv:2010.08895}{}{}{}.
\PrintBackRefs{\CurrentBib}

\bibitem [\protect \citeauthoryear {%
Mellor%
}{%
Mellor%
}{%
{\protect \APACyear {1998}}%
}]{%
mellorUsersGuideThree1998}
\APACinsertmetastar {%
mellorUsersGuideThree1998}%
\begin{APACrefauthors}%
Mellor, G\BPBI L.%
\end{APACrefauthors}%
\unskip\
\newblock
\APACrefYear{1998}.
\newblock
\APACrefbtitle {Users Guide for a Three Dimensional, Primitive Equation,
  Numerical Ocean Model} {Users guide for a three dimensional, primitive
  equation, numerical ocean model}.
\newblock
\APACaddressPublisher{}{{Program in Atmospheric and Oceanic Sciences, Princeton
  University Princeton, NJ}}.
\PrintBackRefs{\CurrentBib}

\bibitem [\protect \citeauthoryear {%
Minuzzi%
\ \BBA {} Farina%
}{%
Minuzzi%
\ \BBA {} Farina%
}{%
{\protect \APACyear {2023}}%
}]{%
minuzziDeepLearningApproach2023}
\APACinsertmetastar {%
minuzziDeepLearningApproach2023}%
\begin{APACrefauthors}%
Minuzzi, F\BPBI C.%
\BCBT {}\ \BBA {} Farina, L.%
\end{APACrefauthors}%
\unskip\
\newblock
\APACrefYearMonthDay{2023}{{\APACmonth{02}}}{}.
\newblock
{\BBOQ}\APACrefatitle {A Deep Learning Approach to Predict Significant Wave
  Height Using Long Short-Term Memory} {A deep learning approach to predict
  significant wave height using long short-term memory}.{\BBCQ}
\newblock
\APACjournalVolNumPages{Ocean Modelling}{181}{}{102151}.
\newblock
\begin{APACrefDOI} \doi{10.1016/j.ocemod.2022.102151} \end{APACrefDOI}
\PrintBackRefs{\CurrentBib}

\bibitem [\protect \citeauthoryear {%
{National Data Buoy Center}%
}{%
{National Data Buoy Center}%
}{%
{\protect \APACyear {2024}}%
}]{%
ndbcData2024}
\APACinsertmetastar {%
ndbcData2024}%
\begin{APACrefauthors}%
{National Data Buoy Center}.%
\end{APACrefauthors}%
\unskip\
\newblock
\APACrefYearMonthDay{2024}{}{}.
\newblock
\APACrefbtitle {{National Data Buoy Center [Dataset]}.} {{National Data Buoy
  Center [Dataset]}.}
\newblock
\APACaddressPublisher{}{{National Oceanic and Atmospheric Administration}}.
\newblock
\begin{APACrefURL} \url{https://www.ndbc.noaa.gov/} \end{APACrefURL}
\newblock
\APACrefnote{Accessed on September 18, 2025}
\PrintBackRefs{\CurrentBib}

\bibitem [\protect \citeauthoryear {%
Pathak%
\ \protect \BOthers {.}}{%
Pathak%
\ \protect \BOthers {.}}{%
{\protect \APACyear {2022}}%
}]{%
pathakFourcastnetGlobalDatadriven2022}
\APACinsertmetastar {%
pathakFourcastnetGlobalDatadriven2022}%
\begin{APACrefauthors}%
Pathak, J.%
, Subramanian, S.%
, Harrington, P.%
, Raja, S.%
, Chattopadhyay, A.%
, Mardani, M.%
\BDBL {}others%
\end{APACrefauthors}%
\unskip\
\newblock
\APACrefYearMonthDay{2022}{}{}.
\newblock
{\BBOQ}\APACrefatitle {Fourcastnet: {{A}} Global Data-Driven High-Resolution
  Weather Model Using Adaptive Fourier Neural Operators} {Fourcastnet: {{A}}
  global data-driven high-resolution weather model using adaptive fourier
  neural operators}.{\BBCQ}
\newblock
\APACjournalVolNumPages{arXiv preprint arXiv:2202.11214}{}{}{}.
\PrintBackRefs{\CurrentBib}

\bibitem [\protect \citeauthoryear {%
Rasp%
\ \protect \BOthers {.}}{%
Rasp%
\ \protect \BOthers {.}}{%
{\protect \APACyear {2020}}%
}]{%
raspWeatherBenchBenchmarkData2020a}
\APACinsertmetastar {%
raspWeatherBenchBenchmarkData2020a}%
\begin{APACrefauthors}%
Rasp, S.%
, Dueben, P\BPBI D.%
, Scher, S.%
, Weyn, J\BPBI A.%
, Mouatadid, S.%
\BCBL {}\ \BBA {} Thuerey, N.%
\end{APACrefauthors}%
\unskip\
\newblock
\APACrefYearMonthDay{2020}{}{}.
\newblock
{\BBOQ}\APACrefatitle {{{WeatherBench}}: A Benchmark Data Set for Data-driven
  Weather Forecasting} {{{WeatherBench}}: A benchmark data set for data-driven
  weather forecasting}.{\BBCQ}
\newblock
\APACjournalVolNumPages{Journal of Advances in Modeling Earth
  Systems}{12}{11}{e2020MS002203}.
\PrintBackRefs{\CurrentBib}

\bibitem [\protect \citeauthoryear {%
Skamarock%
\ \protect \BOthers {.}}{%
Skamarock%
\ \protect \BOthers {.}}{%
{\protect \APACyear {2008}}%
}]{%
skamarockDescriptionAdvancedResearch2008}
\APACinsertmetastar {%
skamarockDescriptionAdvancedResearch2008}%
\begin{APACrefauthors}%
Skamarock, W\BPBI C.%
, Klemp, J\BPBI B.%
, Dudhia, J.%
, Gill, D\BPBI O.%
, Barker, D\BPBI M.%
, Duda, M\BPBI G.%
\BDBL {}Powers, J\BPBI G.%
\end{APACrefauthors}%
\unskip\
\newblock
\APACrefYearMonthDay{2008}{}{}.
\newblock
{\BBOQ}\APACrefatitle {A Description of the Advanced Research {{WRF}} Version
  3} {A description of the advanced research {{WRF}} version 3}.{\BBCQ}
\newblock
\APACjournalVolNumPages{NCAR technical note}{475}{}{113}.
\PrintBackRefs{\CurrentBib}

\bibitem [\protect \citeauthoryear {%
Slivinski%
, Whitaker%
, Frolov%
, Smith%
\BCBL {}\ \BBA {} Agarwal%
}{%
Slivinski%
\ \protect \BOthers {.}}{%
{\protect \APACyear {2025}}%
}]{%
slivinski2025assimilating}
\APACinsertmetastar {%
slivinski2025assimilating}%
\begin{APACrefauthors}%
Slivinski, L\BPBI C.%
, Whitaker, J\BPBI S.%
, Frolov, S.%
, Smith, T\BPBI A.%
\BCBL {}\ \BBA {} Agarwal, N.%
\end{APACrefauthors}%
\unskip\
\newblock
\APACrefYearMonthDay{2025}{}{}.
\newblock
{\BBOQ}\APACrefatitle {Assimilating observed surface pressure into ML weather
  prediction models} {Assimilating observed surface pressure into ml weather
  prediction models}.{\BBCQ}
\newblock
\APACjournalVolNumPages{Geophysical Research Letters}{52}{6}{e2024GL114396}.
\PrintBackRefs{\CurrentBib}

\bibitem [\protect \citeauthoryear {%
Tian%
, Holdaway%
\BCBL {}\ \BBA {} Kleist%
}{%
Tian%
\ \protect \BOthers {.}}{%
{\protect \APACyear {2024}}%
}]{%
tian2024exploring}
\APACinsertmetastar {%
tian2024exploring}%
\begin{APACrefauthors}%
Tian, X.%
, Holdaway, D.%
\BCBL {}\ \BBA {} Kleist, D.%
\end{APACrefauthors}%
\unskip\
\newblock
\APACrefYearMonthDay{2024}{}{}.
\newblock
{\BBOQ}\APACrefatitle {Exploring the use of machine learning weather models in
  data assimilation} {Exploring the use of machine learning weather models in
  data assimilation}.{\BBCQ}
\newblock
\APACjournalVolNumPages{arXiv preprint arXiv:2411.14677}{}{}{}.
\PrintBackRefs{\CurrentBib}

\bibitem [\protect \citeauthoryear {%
{TIGGE Project Archive}%
}{%
{TIGGE Project Archive}%
}{%
{\protect \APACyear {2010}}%
}]{%
tiggeDataset2010}
\APACinsertmetastar {%
tiggeDataset2010}%
\begin{APACrefauthors}%
{TIGGE Project Archive}.%
\end{APACrefauthors}%
\unskip\
\newblock
\APACrefYearMonthDay{2010}{}{}.
\newblock
\APACrefbtitle {{The THORPEX Interactive Grand Global Ensemble (TIGGE) dataset
  [Dataset]}.} {{The THORPEX Interactive Grand Global Ensemble (TIGGE) dataset
  [Dataset]}.}
\newblock
\APACaddressPublisher{}{{European Centre for Medium-Range Weather Forecasts
  (ECMWF)}}.
\newblock
\begin{APACrefURL} \url{https://confluence.ecmwf.int/display/TIGGE}
  \end{APACrefURL}
\newblock
\APACrefnote{As of June 2024, ECMWF is migrating this dataset to a new service
  and the original URL may be inactive}
\PrintBackRefs{\CurrentBib}

\bibitem [\protect \citeauthoryear {%
Tolman%
}{%
Tolman%
}{%
{\protect \APACyear {2009}}%
}]{%
tolmanUserManualSystem2009}
\APACinsertmetastar {%
tolmanUserManualSystem2009}%
\begin{APACrefauthors}%
Tolman, H.%
\end{APACrefauthors}%
\unskip\
\newblock
\APACrefYearMonthDay{2009}{}{}.
\newblock
{\BBOQ}\APACrefatitle {User Manual and System Documentation of {{WAVEWATCH III
  TM}} Version 3.14} {User manual and system documentation of {{WAVEWATCH III
  TM}} version 3.14}.{\BBCQ}
\newblock
\APACjournalVolNumPages{Technical note, MMAB Contribution}{276}{220}{}.
\PrintBackRefs{\CurrentBib}

\bibitem [\protect \citeauthoryear {%
Wiegel%
}{%
Wiegel%
}{%
{\protect \APACyear {1960}}%
}]{%
wiegelWindWavesSwell1960}
\APACinsertmetastar {%
wiegelWindWavesSwell1960}%
\begin{APACrefauthors}%
Wiegel, R\BPBI L.%
\end{APACrefauthors}%
\unskip\
\newblock
\APACrefYearMonthDay{1960}{}{}.
\newblock
{\BBOQ}\APACrefatitle {Wind Waves and Swell} {Wind waves and swell}.{\BBCQ}
\newblock
\APACjournalVolNumPages{Coastal Engineering Proceedings}{}{7}{1--1}.
\PrintBackRefs{\CurrentBib}

\bibitem [\protect \citeauthoryear {%
Xiong%
\ \protect \BOthers {.}}{%
Xiong%
\ \protect \BOthers {.}}{%
{\protect \APACyear {2023}}%
}]{%
xiongAIGOMSLargeAIDriven2023}
\APACinsertmetastar {%
xiongAIGOMSLargeAIDriven2023}%
\begin{APACrefauthors}%
Xiong, W.%
, Xiang, Y.%
, Wu, H.%
, Zhou, S.%
, Sun, Y.%
, Ma, M.%
\BCBL {}\ \BBA {} Huang, X.%
\end{APACrefauthors}%
\unskip\
\newblock
\APACrefYearMonthDay{2023}{}{}.
\newblock
{\BBOQ}\APACrefatitle {{{AI-GOMS}}: {{Large AI-Driven Global Ocean Modeling
  System}}} {{{AI-GOMS}}: {{Large AI-Driven Global Ocean Modeling
  System}}}.{\BBCQ}
\newblock
\APACjournalVolNumPages{arXiv preprint arXiv:2308.03152}{}{}{}.
\PrintBackRefs{\CurrentBib}

\bibitem [\protect \citeauthoryear {%
Zhang%
\ \protect \BOthers {.}}{%
Zhang%
\ \protect \BOthers {.}}{%
{\protect \APACyear {2025}}%
}]{%
Zhang2025zenodo}
\APACinsertmetastar {%
Zhang2025zenodo}%
\begin{APACrefauthors}%
Zhang, Z.%
, Yu, H.%
, Ren, D.%
, Zhang, C.%
, Sun, M.%
\BCBL {}\ \BBA {} Qi, X.%
\end{APACrefauthors}%
\unskip\
\newblock
\APACrefYearMonthDay{2025}{}{}.
\newblock
\APACrefbtitle {{manuscript code parameter and data [Dataset, Software]}.}
  {{manuscript code parameter and data [Dataset, Software]}.}
\newblock
\APACaddressPublisher{}{Zenodo}.
\newblock
\begin{APACrefURL} \url{https://doi.org/10.5281/zenodo.15620954}
  \end{APACrefURL}
\newblock
\begin{APACrefDOI} \doi{10.5281/zenodo.15620954} \end{APACrefDOI}
\PrintBackRefs{\CurrentBib}

\bibitem [\protect \citeauthoryear {%
Zheng%
\ \protect \BOthers {.}}{%
Zheng%
\ \protect \BOthers {.}}{%
{\protect \APACyear {2023}}%
}]{%
zhengMultivariateDataDecomposition2023}
\APACinsertmetastar {%
zhengMultivariateDataDecomposition2023}%
\begin{APACrefauthors}%
Zheng, Z.%
, Ali, M.%
, Jamei, M.%
, Xiang, Y.%
, Abdulla, S.%
, Yaseen, Z\BPBI M.%
\BCBL {}\ \BBA {} Farooque, A\BPBI A.%
\end{APACrefauthors}%
\unskip\
\newblock
\APACrefYearMonthDay{2023}{}{}.
\newblock
{\BBOQ}\APACrefatitle {Multivariate Data Decomposition Based Deep Learning
  Approach to Forecast One-Day Ahead Significant Wave Height for Ocean Energy
  Generation} {Multivariate data decomposition based deep learning approach to
  forecast one-day ahead significant wave height for ocean energy
  generation}.{\BBCQ}
\newblock
\APACjournalVolNumPages{Renewable and Sustainable Energy
  Reviews}{185}{}{113645}.
\PrintBackRefs{\CurrentBib}

\end{thebibliography}

%%%%%%%%%%%%%%%%%%%%%%%%%%%%%%%%%%%%%%%%%%%%%%%

%\bibliography{ enter your bibtex bibliography filename here }

%Reference citation instructions and examples:
%
% Please use ONLY \cite and \citeA for reference citations.
% \cite for parenthetical references
% ...as shown in recent studies (Simpson et al., 2019)
% \citeA for in-text citations
% ...Simpson et al. (2019) have shown...
%
%
%...as shown by \citeA{jskilby}.
%...as shown by \citeA{lewin76}, \citeA{carson86}, \citeA{bartoldy02}, and \citeA{rinaldi03}.
%...has been shown \cite{jskilbye}.
%...has been shown \cite{lewin76,carson86,bartoldy02,rinaldi03}.
%... \cite <i.e.>[]{lewin76,carson86,bartoldy02,rinaldi03}.
%...has been shown by \cite <e.g.,>[and others]{lewin76}.
%
% apacite uses < > for prenotes and [ ] for postnotes
% DO NOT use other cite commands (e.g., \citet, \citep, \citeyear, \nocite, \citealp, etc.).
%

\end{document}